\documentclass[journal, hidelinks]{IEEEtran}

\usepackage{amsmath}
\usepackage{amssymb}
\usepackage{amsfonts}
\usepackage{amsthm}
\usepackage{bbm}
\usepackage{mathtools}
\usepackage{physics}
\usepackage[utf8]{inputenc}
\usepackage[T1]{fontenc}
\usepackage{textcomp}
\usepackage{graphicx}

\usepackage{nicefrac}
\usepackage{booktabs}
\usepackage{multirow}
\usepackage{xcolor}
\usepackage{hyperref}
\newtheorem{proposition}{Proposition}

\newcommand{\inner}[2]{\ensuremath{\langle #1,#2 \rangle}}

\DeclareMathOperator*{\argmin}{arg\,min}

\let\norm\undefined % <-- "Undefine" \norm
\DeclarePairedDelimiter\norm{\lVert}{\rVert}

\newcommand{\R}{\ensuremath{\mathbb{R}}}

\newcommand{\J}{\ensuremath{\mathcal{J}}}
\newcommand{\X}{\ensuremath{\mathcal{X}}}
\newcommand{\Y}{\ensuremath{\mathcal{Y}}}
\newcommand{\E}{\ensuremath{\mathbb{E}}}
\newcommand{\expect}[2][]{\ensuremath{\E_{#1}\left[#2\right]}}
\newcommand{\A}{\ensuremath{\mathbf{A}}}
\newcommand{\Rec}{\ensuremath{\mathbf{R}}}

\newcommand{\ind}[1][\cdot]{\ensuremath{\mathbbm{1}_{#1}}}

\newcommand{\absorption}{\ensuremath{\alpha}}
\newcommand{\numsplits}{\ensuremath{K}}
\newcommand{\numangles}{N_\theta}
\newcommand{\numpixelsy}{N_p}
\newcommand{\numpixelsx}{N_x}
\renewcommand{\angle}{\ensuremath{\theta}}
\renewcommand{\vec}[1]{\mathbf{#1}}
\newcommand{\rv}[1]{\mathsf{#1}}
\newcommand{\instance}[1]{\mathrm{#1}}
\newcommand{\funparam}[1]{\mathit{#1}}
\newcommand{\x}{\vec{x}}

\newcommand{\xrv}{\rv{x}}
\newcommand{\xinst}{\instance{x}}
\newcommand{\xpar}{\funparam{x}}
\newcommand{\rclean}{\x^{*}}
\newcommand{\rcleanrv}{\rv{x^*}}

\newcommand{\rcleanJ}{\rcleanrv_{\jrv}}
\newcommand{\rcleaninst}{\instance{x}^{*}}
\newcommand{\rnoisy}{\ensuremath{\tilde{\x}}}
\newcommand{\rnoisyrv}{\rv{\tilde{x}}}
\newcommand{\rnoisyinst}{\instance{\tilde{{x}}}}
\newcommand{\rnoisypar}{\funparam{\tilde{x}}}
\newcommand{\rnoisyC}{\rnoisyrv_{\jrv^C}}
\newcommand{\rnoisyJ}{\rnoisyrv_{\jrv}}
\newcommand{\pclean}{\vec{y}}
\newcommand{\pcleanrv}{\rv{y}}
\newcommand{\pcleanvec}{\vec{y}}
\newcommand{\pcleaninst}{\instance{y}}
\newcommand{\pcleanpar}{\funparam{y}}
\newcommand{\pnoisy}{\vec{\tilde{y}}}

\newcommand{\pnoisyrv}{\rv{\tilde{y}}}
\newcommand{\pnoisypar}{\funparam{\tilde{y}}}
\newcommand{\pnoisyinst}{\instance{\tilde{y}}}
\newcommand{\pnoisyinstance}{\pnoisyinst}
\renewcommand{\j}{J}
\newcommand{\jrv}{\rv{\j}}

\newcommand{\jpar}{\funparam{\j}}

\newcommand{\noise}{\epsilon}

\newcommand{\param}{\ensuremath{\varphi}}
\newcommand{\paramhat}{\ensuremath{\hat \varphi}}

\newcommand{\network}{\ensuremath{f_{\param}}}
\newcommand{\trainednetwork}{\ensuremath{f_{\paramhat}}}

\usepackage{cite}
\ifCLASSINFOpdf
\else
\fi
\usepackage{array}
\ifCLASSOPTIONcompsoc
 \usepackage[caption=false,font=normalsize,labelfont=sf,textfont=sf]{subfig}
\else
 \usepackage[caption=false,font=footnotesize]{subfig}
\fi
\usepackage{fixltx2e}
\begin{document}

\title{
  Noise2Inverse: Self-supervised deep convolutional denoising for
    tomography
}


\author{
  Allard~A.~Hendriksen,
  Dani\"el~M.~Pelt,
  K.~Joost~Batenburg,~\IEEEmembership{Member,~IEEE}%
  \thanks{
    A. A. Hendriksen, D. M. Pelt, and  K. J. Batenburg are with the
    Centrum Wiskunde \& Informatica, Amsterdam 1098 XG, The
    Netherlands
    (e-mail: allard.hendriksen@cwi.nl;d.m.pelt@cwi.nl;joost.batenburg@cwi.nl).}%
  \thanks{
    K. J. Batenburg is with the
    Leiden Institute of Advanced Computer Science, Leiden Universiteit, 2333 CA Leiden, The
    Netherlands}%
  \thanks{
    %Manuscript received April 19, 2005; revised August 26, 2015.
  }}


\markboth{IEEE TRANSACTIONS ON COMPUTATIONAL IMAGING,~Vol.~X, No.~Y, Month~Year}%
{TODO \MakeLowercase{\textit{et al.}}: TODO}



\maketitle

\begin{abstract}
  Recovering a high-quality image from noisy indirect measurements is
  an important problem with many applications.
  For such inverse problems, supervised deep convolutional neural
  network (CNN)-based denoising methods have shown strong results, but
    the success of these supervised methods
  critically depends on the availability of a
  high-quality training dataset of similar measurements.
  For image denoising, methods are available that enable training
  without a separate training dataset by assuming that the noise in
  two different pixels is uncorrelated.
  However, this assumption does not hold for inverse problems,
  resulting in artifacts in the denoised images produced by existing methods.
  Here, we propose Noise2Inverse, a deep CNN-based denoising method
    for linear image reconstruction algorithms
  that does not require any
  additional clean or noisy data.
  Training a CNN-based denoiser is enabled by exploiting the noise
  model to compute multiple statistically independent reconstructions.
  We develop a theoretical framework which shows that such training
  indeed obtains a denoising CNN, assuming the measured noise is
  element-wise independent and zero-mean.
    On simulated CT datasets, Noise2Inverse demonstrates
    an  improvement in peak signal-to-noise ratio and
    structural similarity index compared to state-of-the-art image
    denoising methods and conventional reconstruction methods, such as
    Total-Variation Minimization.
  We also demonstrate that the method is able to significantly reduce
  noise in challenging real-world experimental datasets.
\end{abstract}


\IEEEpeerreviewmaketitle{}

\section{Introduction}

\IEEEPARstart{R}{econstruction}
  algorithms compute an image from
  indirect measurements.
  For a subclass of these algorithms, the relation between the
  reconstructed image and the measured data can be described by a linear
  operator.
  Such \emph{linear reconstruction methods} are used in a variety of
  applications, including X-ray and photo-acoustic tomography, ultrasound
  imaging, deconvolution microscopy, and X-ray
  holography~\cite{sibarita-2005-decon-micros,marone-2012-regrid-recon,
    mccollough-2017-low-dose,matrone-2015-delay-multip,
    kuchment-2011-mathem-photoac,poudel-2019-survey-comput,
    zabler-2005-optim-phase,artioli-2010-x-ray,zeng-2019-analy-time}.
  These methods are well-suited for fast, parallel
  computation~\cite{pelt-2018-improv-tomog}, but are also generally sensitive to
  measurement noise, leading to errors in the reconstructed
  image~\cite{buzug-2008-comput-tomog,sibarita-2005-decon-micros}.
Controlling this error, i.e., \emph{denoising}, is a central problem
in inverse problems in imaging~\cite{belthangady-2019-applic-promis, kang-2017-deep-convol,
  mccollough-2017-low-dose, pelt-2018-improv-tomog,
  sun-2018-effic-accur, chang-2013-asses-model,
  buchholz-2019-cryo-care}.

  Supervised
deep convolutional neural network (CNN)-based methods are able to
accurately denoise reconstructed images in several
inverse problems~\cite{belthangady-2019-applic-promis,
  kang-2017-deep-convol, mccollough-2017-low-dose,
  pelt-2018-improv-tomog, sun-2018-effic-accur}.
These networks are trained in a \emph{supervised} setting,
which amounts to finding the network parameters that best compute a
mapping from noisy to clean reconstructed images on a dataset of
example image pairs.
However, the success of these
  supervised
deep learning methods critically depends
on the availability of such a high-quality training dataset of
similar images~\cite{belthangady-2019-applic-promis,liu-2020-rare}.

For photographic image denoising, recent work has shown that deep
learning may be possible without obtaining high quality target images,
by instead training on paired noisy images~\cite{lehtinen-2018-noise2}.
Nonetheless, such \emph{Noise2Noise} training still requires
additional noisy data.
The feasibility of image denoising by \emph{self-supervised} training,
that is, training with \emph{single} instead of paired noisy
images, was demonstrated by~\cite{krull-2018-noise2-learn,batson-2019-noise2,laine-2019-high-qualit}.
These self-supervised training methods, such as Noise2Self, depend on
the assumption that noise in one pixel is statistically independent
from noise in another pixel.

In inverse problems, reconstructed images may exhibit coupling of the
measured noise~\cite{kang-2017-deep-convol}.
In CT, for instance, back-projection smears out the noise in a detector
pixel across a line through the reconstructed image.
Naturally, this causes the noise in one pixel to be statistically
dependent on noise in other pixels of the reconstructed image.

In this paper, we demonstrate that a straightforward application of
Noise2Self to reconstructed CT images delivers substantially inferior
results compared to results obtained on photographic images, for which
it was developed.
We analyze the cause of this apparent mismatch, and propose
Noise2Inverse, a new approach that is specifically designed for linear
reconstruction methods in imaging to overcome these limitations.

In the proposed Noise2Inverse approach, the training regime explicitly
takes into account the structure of the noise in the inverse problem.
  In its simplest form, our method splits the measured data in two
  parts, from which two reconstructions are computed.
  We train a CNN to transform one reconstruction into the other, and
  vice versa.
  The properties of the physical forward model cause the noise in the
  reconstructed images to be statistically independent.
This enables the CNN to perform \emph{blind} image denoising on the
reconstructed images.
  That is, our method does not assume a \emph{known noise model}.
We stress that our method can be applied to existing datasets without
acquiring additional data.

  In recent years, a range of deep learning approaches have been developed for
  denoising in imaging with limited training data.
  Several weight-regularized self-supervised methods exist that require a known
  Gaussian noise
  model~\cite{soltanayev-2018-train,zhussip-2019-exten-stein,metzler-2018-unsup-learn,cha-2019-boost-cnn}.
  While such a model is often available in direct imaging modalities, the noise
  model for reconstructed images in an inverse problem setting is often more
  complex and hard to characterize by such a Gaussian model.
  Unsupervised approaches using the Deep Image
  Prior~\cite{ulyanov-2017-deep-image-prior,cheng-2019-bayes-persp,mataev-2019-deepr}
  have been proposed for image restoration and inverse
  problems~\cite{jin-2019-time-depen,dittmer-2018-regul-by-archit}.
  A key obstacle for the application of such techniques to large-scale 3D
  image reconstruction problems is their computational cost, as they involve
  training a new network for every 2D slice of the reconstruction. For inverse
  problems, approaches that rely on splitting the measurement data have
  recently been proposed for magnetic resonance imaging
  (MRI)~\cite{liu-2020-rare,yaman-2020-self-super} and Cryo-transmission
  electron microscopy (Cryo-EM)~\cite{buchholz-2019-cryo-care} showing image
  quality improvement with respect to denoising applied on the reconstructed
  image. While these results are highly promising, a solid theoretical
  underpinning that allows analysis and insights into the interplay between
  the underlying noise model of the inverse problem and the obtained solution
  is currently lacking.

  In this paper --- motivated by these promising results --- we present a
  framework for generalizing the self-supervised denoising approach in the
  setting of linear reconstruction methods.
  Our framework pinpoints exactly the underlying theoretical properties that
  explain the differences in observed results of self-supervised approaches.
  We perform a qualitative and quantitative comparison to conventional iterative
  reconstruction and state-of-the-art image denoising techniques.
  We evaluate these methods on several simulated low-dose CT datasets, and
  include results on an existing experimentally acquired CT dataset, for which
  no low-noise data is available.
  In addition, we present a systematic analysis of the hyper-parameters of the
  proposed method.

This paper is structured as follows.
In Section~\ref{sec:notation-concepts}, we introduce linear inverse
problems and deep learning for image denoising, including
self-supervised methods.
In Section~\ref{sec:method}, we introduce the proposed Noise2Inverse
method, and show its theoretical properties, which we use to develop
an implementation for computed tomography.
In Section~\ref{sec:results}, we perform experiments to compare the
performance of Noise2Inverse, conventional reconstruction techniques,
and Noise2Self-based methods on real and simulated CT datasets.
In addition, we perform a hyper-parameter study of the proposed
method.
We discuss these results in Section~\ref{sec:discussion}, and conclude
in Section~\ref{sec:conclusion}.

\section{Notation and concepts}\label{sec:notation-concepts}

As prerequisites for describing our Noise2Inverse approach, we first
discuss deep learning methods for image denoising, including
strategies for training neural networks when clean images are
unavailable.
In addition, we review linear inverse problems, where we discuss that
denoising reconstructed images introduces additional difficulties.

\subsection{Deep learning for image denoising}\label{sec:deep-learning}

The goal of image denoising is to recover a 2D image $\pclean \in \Y=\R^m$
from a measurement $\pnoisy \in \Y$ that is corrupted by random
noise \(\noise\), taking values in \(\Y\).
This problem is described by the equation
\begin{align}
\label{eq:image-denoising}
\pnoisy = \pclean + \noise.
\end{align}
It is common to assume that the entries of the noise vector
\(\noise\) are mutually independent.
Many image denoising methods rely on this assumption~\cite{zhang-2017-beyon-gauss-denois,dabov-2007-image-denois,krull-2018-noise2-learn}.
In addition, these methods assume that the image exhibits some
statistically meaningful structure that can be exploited to remove the
noise.
The popular BM3D algorithm~\cite{dabov-2007-image-denois}, for example,
exploits non-local self-similarity, i.e., the expectation that
certain structures of the image are repeated elsewhere in the image.
  Note that it is also possible to include BM3D as a prior inside iterative
  algorithms for inverse problems using a plug-and-play
  framework~\cite{venkatakrishnan-2013-plug-and}.

Instead of relying on an explicit image prior, prior knowledge can be
based on a range of example images, as is done in deep learning.
In particular, deep convolutional neural networks (CNNs) have been
recognized as a powerful and versatile denoising technique~\cite{zhang-2017-beyon-gauss-denois}.
We briefly introduce three training schemes for denoising with CNNs:
supervised\cite{zhang-2017-beyon-gauss-denois}, Noise2Noise~\cite{lehtinen-2018-noise2}, and Noise2Self~\cite{batson-2019-noise2}.

\begin{figure*}[]
  \centering
  \includegraphics[]{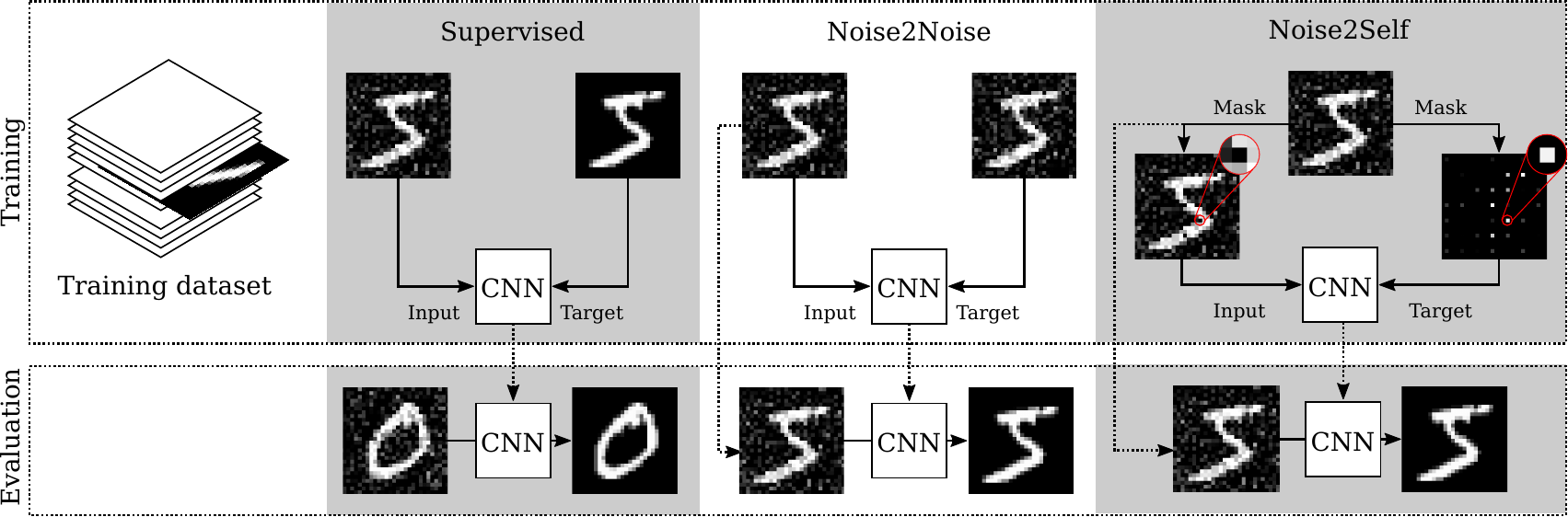}
  \caption[]{
    Three training regimes for CNN-based image denoising.
    Supervised training is performed with noisy and clean images, and
    the trained CNN is applied to unseen noisy data.
    Noise2Noise training is performed with pairs of noisy images.
    Noise2Self training is performed with just noisy images, which are
    split into input-target pairs.
    The loss is only computed where target pixels are non-zero.
    The red inset displays one of these locations.
    For Noise2Noise and Noise2Self, the trained CNN can be applied to the
    training data to obtain clean images.
  }\label{fig:training-schemes}
\end{figure*}

The \textbf{supervised} training scheme has access to a \emph{training dataset}
containing pairs of noisy \emph{input} and clean \emph{target} images
\begin{align}
(\pnoisyinst_i, \pcleaninst_i) \sim (\pcleanrv + \noise, \pcleanrv), \quad i=1, \ldots, N,
\end{align}
where \(\pcleanrv\) is a random variable taking values in \(\Y\) that
represents the clean images.
  The supervised training objective is to find the \emph{regression}
  function
\begin{align}
\label{eq:supervised-objective}
h^{*} = \argmin_h
\expect[\pcleanrv, \noise]{\norm*{h(\pcleanrv + \noise) - \pcleanrv}_2^2},
\end{align}
  that minimizes the \emph{expected prediction error}~\cite{hastie-2009-elemen-statis-learn}.
  The most common loss function is the pixel-wise mean square error,
  which we use here.
  Alternative training losses are also used, such as the L1 loss and perceptual
  losses~\cite{lehtinen-2018-noise2}.
Solving Equation~\eqref{eq:supervised-objective} is usually
intractable.
Therefore, the expectation is estimated by the sample mean over the
training dataset, which is minimized over neural networks
\(\network:\Y \to \Y\) with parameters \(\param\).
The training task is then to find the optimal parameters
\begin{align}
\label{eq:supervised-training-task}
\paramhat = \argmin_{\param} \sum_{i=1}^N \norm*{\network(\pnoisyinst_i) - \pcleaninst_i}_2^2,
\end{align}
which minimize the loss on the sampled image pairs.
The trained network $\trainednetwork$ is applied to unseen noisy
images to obtain denoised images, as displayed in Figure~\ref{fig:training-schemes}.

  The regression function that minimizes the expected
  prediction error in Equation~\eqref{eq:supervised-objective} is the
  conditional expectation
\begin{align}
\label{eq:supervised-minimizer}
h^*(\pnoisypar) = \expect{\pcleanrv \mid \pcleanrv + \noise = \pnoisypar}.
\end{align}
In practice, the trained neural network \(\trainednetwork\) does not
equal \(h^{*}\) and an approximation is obtained.

\textbf{Noise2Noise} training may be applied if no clean images are
available, but one can measure \emph{independent} instances of the
noise for each image.
The training dataset contains pairs of independent noisy images
\begin{align}
(\pcleaninst_i + \noise_i, \pcleaninst_i + \delta_i) \sim (\pcleanrv + \noise, \pcleanrv + \delta), \quad i=1, \ldots, N,
\end{align}
where the noise \(\delta\) is a random variable that is statistically
independent of \(\noise\).
The training task is to determine
\begin{align}
\label{eq:noise2noise-training-task}
\paramhat = \argmin_{\param} \sum_{i=1}^N \norm*{\network(\pcleaninst_i + \noise_i) - (\pcleaninst_i + \delta_i)}_2^2,
\end{align}
and the trained neural network \(\trainednetwork\) approximates
\begin{align}
\label{eq:noise2noise-objective}
h^{*} = \argmin_h
\expect[\pcleanrv, \noise, \delta]
      {\norm*{h(\pcleanrv + \noise) - (\pcleanrv + \delta)}_2^2}.
\end{align}
If the noise \(\delta\) is mean-zero, i.e.,
\(\expect{\delta} = 0\),
  the expected prediction error in
Equation~\eqref{eq:noise2noise-objective} is
minimized by the same
  regression
function \(h^{*}\) as in the supervised regime
(Equation~\eqref{eq:supervised-minimizer}).
In practice, Noise2Noise and supervised training indeed yield
trained networks with similar denoising performance.

\textbf{Noise2Self} enables training a neural network denoiser without any
additional images.
The training dataset contains only noisy images
\begin{align}
\pnoisyinst_i \sim \pcleanrv + \noise, \quad i=1, \ldots, N.
\end{align}
The method depends on the assumption that the noise is element-wise
statistically independent and mean-zero, and that the clean images
exhibit some spatial correlation.

Noise2Self training uses a masking scheme that ensures that the loss
compares two statistically independent images.
For simplicity, we describe a simplified version of Noise2Self
training, and refer to~\cite{batson-2019-noise2} for a more in-depth
explanation.
In each training step, the noisy image is
  split
into two
  \emph{sub-images}: one sub-image --- the target --- contains
  non-adjacent pixels and the other sub-image --- the input ---
  contains the remaining surrounding pixels.
The network is trained to predict the value of a noisy pixel
from its surrounding noisy pixels,
  as is shown in Figure~\ref{fig:training-schemes}.

  The division of pixels between the input and target image is
  determined by a partition \(\J\) of the pixels such that adjacent
  pixels are in different subsets.  We denote by $J \in \J$ the
  \emph{target section}, and by $J^C$ the \emph{input section}, where
  \(J^C\) denotes the set complement of \(J\), containing all pixel
  locations not contained in \(J\).  The input and target images
  $\ind[J^C]\pnoisyinst_i$ and $\ind[J]\pnoisyinst_i$ have non-zero
  pixels only in the \emph{input} and \emph{target section},
  respectively.  Here, $\ind[J]$ denotes the indicator function such
  that element-wise multiplication of \(\ind[J]\) with an image
  retains pixel values in \(J\) and sets pixels to zero elsewhere.
  The training task is to determine the set of network parameters
  minimizing the training loss
\begin{align}
\label{eq:noise2self-finite-sample-training-task}
\paramhat = \argmin_{\param} \sum_{i=1}^{N} \sum_{J \in \J} \norm{\ind[J] \network(\ind[J^C] \pnoisyinst_i)- \ind[J] \pnoisyinst_i }_2^2,
\end{align}
  where the loss is only computed on the target sections.

  The inference step is performed by the \emph{section-wise combined
    network} \(g_{\paramhat}:\Y\to\Y\),
  \begin{align}
    \label{eq:noise2self-j-invariant}
    g_{\paramhat}(\pnoisypar) := \sum_{J \in \J} \ind[J] \trainednetwork(\ind[J^C] \pnoisypar ),
  \end{align}
  that computes the output in each target section by applying the
  trained network to the input section.

  The piecewise-combined network is an approximation of the regression
  function
  \begin{align}
    \label{eq:noise2self-minimizer}
    g^*(\pnoisypar) = \sum_{J\in\J}\ind[J] \E [ \ind[J] \pcleanrv  \mid \ind[J^C] \left(\pcleanrv + \noise \right) = \ind[J^C] \pnoisypar ].
  \end{align}
  This regression function computes the conditional expectation of the
  clean image in each target section using the surrounding noisy pixels.

Although aforementioned methods can produce accurately denoised
photographic images in many cases~\cite{zhang-2017-beyon-gauss-denois,lehtinen-2018-noise2,batson-2019-noise2},
a subclass of these algorithms --- Noise2Self in particular --- has
strong requirements on the element-wise independence of the noise.
These requirements do not generally hold for solutions of linear
inverse problems, as we discuss next.

\subsection{Linear inverse problems}
We are concerned with inverse problems that are described by the
equation
\begin{align}
\label{eq:inverse-problem}
\A \x = \pclean,
\end{align}
where \(\x \in \X = \R^n\) denotes an unknown image that we wish
to recover, and \(\pclean \in \Y = \R^m\) denotes the indirect
measurement.
The linear forward operator \(\A: \R^n \to \R^m\) describes the physical
model by which the measurement arises from the image \(\x\).
As in the image denoising setting, these measurements are
corrupted by element-wise independent noise \(\noise\), and we write
\begin{align}
\label{eq:noisy-inverse-problems}
\pnoisy &= \A \x + \noise.
\end{align}
  Although noise in Equation~\eqref{eq:noisy-inverse-problems} is
  modeled as an additive term, we note that this model also covers
  non-additive noise, such as Poisson noise, where the noise term
  typically depends on the signal intensity.

Reconstruction algorithms approximate the image \(\x\) from measured
data \(\pclean\).
A subclass of these reconstruction algorithms computes a linear
operator \(\Rec: \Y \to \X\).
Examples of linear reconstruction algorithms include
the filtered backprojection algorithm for tomography and Wiener filtering
for deconvolution microscopy~\cite{buzug-2008-comput-tomog,sibarita-2005-decon-micros}.
We denote the reconstruction from a noisy measurement by
\begin{align}
\label{eq:inverse-problems-reconstructions}
\rnoisy = \Rec \pnoisy = \Rec \pclean + \Rec \noise,
\end{align}
  which can contain artifacts unrelated to the measurement noise,
  e.g., under-sampling artifacts and/or reconstruction artifacts.
The reconstruction operator \(\Rec\) may cause elements of the
reconstructed noise \(\Rec \noise\) to be statistically coupled, even if
\(\noise\) is element-wise independent~\cite{kang-2017-deep-convol}.
That \(\Rec \noise\) does not satisfy the element-wise independence
property is unavoidable for all but the most trivial cases, since
inverse problems are essentially defined by the intricate coupling of
the unknown image with its indirect measurement.

This coupling of the noise seriously degrades the effectiveness of the
Noise2Self approach, as we will see in Section~\ref{sec:results-noise2self}.
In the next section, we propose a self-supervised method that does
take into account the properties of noise in inverse problems.

\section{Noise2Inverse}\label{sec:method}

  In this section, we present the proposed Noise2Inverse method.
  First, we describe the assumed noise model, and give a general description of
  the method.
  In Section~\ref{sec:convergence}, we provide a theoretical explanation how and
  why the convolutional neural network learns to denoise.
  Here, we also discuss how these results can guide implementation in
  practice.
  In Section~\ref{sec:noise2inverse-tomography}, we give a more practical
  description of the implementation for tomography, and discuss implementation
  choices with regard to the obtained theoretical results.

Suppose that we wish to examine several unknown images
$\xinst_1, \ldots, \xinst_N \sim \xrv$, sampled from some random
variable $\xrv$.
We obtain noisy indirect measurements
\begin{align}
\label{eq:4}
  \pnoisyinst_i \sim \A \xinst_i + \noise, \quad i=1, \ldots, N,
\end{align}
where we assume that the noise $\noise$ is element-wise independent
and mean-zero conditional on the data, i.e.,
\begin{align}
\label{eq:noise-zero-mean}
\expect[\xrv, \noise]{\A \xrv + \noise \mid \A \xrv=\pcleaninst} = \pcleaninst.
\end{align}
  As in Equation~\eqref{eq:noisy-inverse-problems}, we assume the noisy may be non-additive.
  Our goal is to recover the \emph{clean reconstructions} that would have
  been obtained in the absence of noise, i.e.,
  $\rcleaninst_i = \Rec \pcleaninst_i$ with
  $\pcleaninst_i = \A \xinst_i, i = 1, \ldots, N$.

One approach is to compute noisy reconstructions, and use Noise2Self
to remove the noise in the reconstructed images.
Given the noisy reconstructions
$\rnoisyinst_i~=~\Rec \pnoisyinst_i, i=1, \ldots, N$, the training
task is to determine
  the network parameters minimizing the training loss
\begin{align}
\label{eq:naive-noise2self-training}
\paramhat = \argmin_{\param}
\sum_{J \in \J_x} \sum_{i=1}^N
    \norm*{\ind[J] \network(\ind[J^C] \rnoisyinst_i ) - \ind[J] \rnoisyinst_i }_2^2,
\end{align}
where
  the target sections are contained in
\(\J_x\), a partition of the pixels of the reconstructed
images.
As discussed before, however, the noise in the
  input and target pixels of the reconstructed images
are unlikely to be statistically independent.

The key idea of the proposed Noise2Inverse method is that it
partitions the data in the measurement domain --- where the noise is
element-wise independent --- but trains the CNN in the reconstruction
domain.
In each training step, the measured data is partitioned into an input
and target component, and a neural network is trained to predict the
reconstruction of one from the reconstruction of the other.
After training, the neural network is applied to denoise the
reconstructions.

  The division of measured data between input and target is determined
  by the collection \(\J\) of \emph{target sections}
  \(J \subset \{1, 2, \ldots, m\}\) that represent subsets of the
  measurement domain \(\Y=\R^m\).
    We note that $\J$ can be chosen such that it contains structured subsets of
    the measurement domain, rather than \emph{all} subsets.
  For each target section $J\in\J$, the measurement is split into
  input and target sub-measurements $\pnoisyinst_{i, J^C}$ and
  $\pnoisyinst_{i, J}$, where \(J^C\) denotes the set complement of
  $J$ with respect to $\{1, 2, \ldots, m\}$.
  The input and target sub-reconstructions are computed by linear
  reconstruction operators \(\Rec_J: \Y_J \to \X\) that take into
  account only the measurements in section $J\in\J$.
  We define
  \begin{align*}
    \rnoisyinst_{i, J^C} = \Rec_{J^C} \pnoisyinst_{i, J^C} \text{ and }
    \rnoisyinst_{i, J} = \Rec_{J} \pnoisyinst_{i, J}
  \end{align*}
  to be the input and target sub-reconstructions of $\pnoisyinst_i$,
  respectively.

  The training task is to determine the parameters
  \begin{align}
    \label{eq:noise2inverse-training-task}
    \paramhat = \argmin_{\param} \frac{1}{\abs{\J}}
    \sum_{J \in \J} \sum_{i=1}^N
    \norm*{\network(\rnoisyinst_{i, J^C} ) - \rnoisyinst_{i, J} }_2^2,
  \end{align}
  that best enable the network \(\trainednetwork\) to predict the
  target sub-reconstruction from the complementary input
  sub-reconstruction.

  The final output is computed by the \emph{section-wise averaged
    network}, which applies the trained network to each input
  sub-reconstruction, and computes the average, yielding
\begin{align}
  \label{eq:1}
  \rcleaninst_{i, \text{out}}
  &= \frac{1}{\abs{\J}} \sum_{J \in \J} \trainednetwork\left(\rnoisyinst_{i, J^C}\right).
  % g_{\paramhat}(\pnoisypar) = \frac{1}{\abs{\J}} \sum_{J \in \J} \trainednetwork(\rnoisypar_{J^C}).
\end{align}
In the next section, we show why the final result approximates the
clean reconstruction.

\subsection{Theoretical framework}\label{sec:convergence}

In this section, we embed Noise2Inverse in a theoretical
framework that explains why it is an accurate denoising method.
In addition, we describe design considerations that enable it to
operate successfully.

  Below, we show that Noise2Inverse recovers an average clean
  reconstruction in theory.
  This result is founded upon Proposition~\ref{proposition:loss-decomposition}, which shows that the expected
  prediction error is the sum of the variance of the reconstructed
  noise and the \emph{supervised prediction error}, which is the
  expected prediction error that would have obtained if the target
  reconstructions were noise-free.
  Hence, the regression function that minimizes the expected
  prediction error also minimizes the loss with respect to the unknown
  clean reconstruction.
  Therefore, it predicts a clean sub-reconstruction when given a noisy
  sub-reconstruction.

  As before, we represent the clean and noisy measurements by the
  random variables \(\pcleanrv = \A \xrv\) and
  \(\pnoisyrv = \pcleanrv + \noise\).  The input and target
  sub-reconstructions are represented by random variables
  $\rnoisyrv_{J^C} = \Rec_{J^C} \pnoisyrv_{J^C}$ and
  $\rnoisyrv_J = \Rec_{J} \pnoisyrv_{J}$ for $J \in \J$.  In this
  case, the trained network \(\trainednetwork\) obtained in Equation~\eqref{eq:noise2inverse-training-task} approximates the regression
  function
  \begin{align}
    \label{eq:empirical-risk-minimization}
    h^{*}
    &= \argmin_h \frac{1}{\abs{\J}}
      \sum_{J \in \J} \E_{\xrv, \noise}
      \norm*{h(\rnoisyrv_{J^C}) - \rnoisyrv_J}^2,
  \end{align}
  which minimizes the expected prediction error.  We randomize the
  section $J$ as well, representing it by \(\jrv\) taking values
  uniformly at random in \(\J\).  The input and target
  sub-reconstructions become random in $\jrv$ as well, which is
  denoted by $\rnoisyC = \Rec_{\jrv^C} \pnoisyrv_{\jrv^C}$ and
  $\rnoisyJ = \Rec_{\jrv} \pnoisyrv_{\jrv}$.  The expected prediction
  error then becomes
  \begin{align*}
    \frac{1}{\abs{\J}}\sum_{J \in \J} \E_{\xrv, \noise}
    \norm*{h(\rnoisyrv_{J^C}) - \rnoisyrv_J}^2
    &= \E_{\mu} \norm*{h(\rnoisyC) - \rnoisyJ}^2,
  \end{align*}
  where we replace the average over \(J \in \J\) by the expectation
  with respect to \(\jrv\).  We denote with \(\mu\) the joint measure
  of \(\xrv, \noise\), and \(\jrv\).  Define the sub-reconstruction of
  the clean measurement
  \begin{align}
    \label{eq:7}
    \rcleanJ &= \Rec_{\jrv} \pcleanrv_{\jrv},
  \end{align}
  which describes the clean target reconstruction.  Now the expected
  prediction error can be decomposed into two parts.
\begin{proposition}[\em Expected prediction error decomposition]\label{proposition:loss-decomposition}
Let \(\rnoisyJ, \rnoisyC, \rcleanJ\), and \(\mu\) be as above.
Let $\noise$ be element-wise independent and satisfy~\eqref{eq:noise-zero-mean}.
Let $\Rec_J$ be linear for all $J \in \J$.
Then, for any measurable function \(h : \X \to \X\), we have
\begin{multline}
\label{eq:loss-decomposition}
\E_{\mu} \norm*{h(\rnoisyC) - \rnoisyJ}_2^2
= \E_{\mu} \norm*{h(\rnoisyC) - \rcleanJ}_2^2 \\
+ \E_{\mu} \norm*{\rcleanJ - \rnoisyJ}_2^2.
\end{multline}
\end{proposition}

  Proposition~\ref{proposition:loss-decomposition} states that the
  expected prediction error can be decomposed into the supervised
  prediction error, which depends on the choice of \(h\), and the
  variance of the reconstruction noise, which does not depend on
  \(h\).  Therefore, when minimizing~\eqref{eq:loss-decomposition},
  the function \(h\) minimizes the difference between its output and
  the unknown clean target sub-reconstruction $\rcleanJ$.
    Note that the minimization of $h$ occurs with respect to $\rcleanJ$ instead
    of the fully sampled reconstruction $\rclean$.
    When the target sections have been chosen such that
    $\expect[\jrv]{\rcleanJ} = \rclean$ holds, however, the difference is
    minimized.

  The supervised prediction error, \(\E_{\mu}\norm{h(\rnoisyC) - \rcleanJ}_2^2\),
  is minimized~\cite{adler-2018-deep-bayes-inver} by the regression function
  \begin{align}
    \label{eq:loss-decomposition-minimizer}
    h^{*}(\rnoisypar)
    &= \expect[\mu]{\rcleanJ \mid \rnoisyC = \rnoisypar}.
  \end{align}

  The section-wise averaged network, defined in Equation~\eqref{eq:1},
  therefore approximates the section-wise average of the regression
  function, defined by
  \begin{align}
    \label{eq:output-is-clean}
    g^*(\pnoisypar) &= \frac{1}{\abs{\J}} \sum_{J \in \J} \expect[\mu]{\rcleanJ \mid \rnoisyC = \rnoisypar_{J^C}},
  \end{align}
  where we write \(\rnoisypar_{J^C} = \Rec_{J^C} \pnoisypar_{J^C}\) for
  $\pnoisypar \in \Y$ and \(J \in
  \J\).

  Using these results, we can explain why the section-wise average obtains a
  denoised output.
  A noisy sub-reconstruction can be explained by different values of
  the clean reconstruction \(\rcleanrv\).
The expectation
$\expect[\mu]{\rcleanJ \mid \rnoisyC = \rnoisypar_{J^C}}$ is the mean
of noiseless reconstructed images consistent with the observed noisy
reconstruction \(\rnoisypar_{J^C}\).
  Equation~\eqref{eq:loss-decomposition-minimizer} therefore predicts
  that our method produces denoised images.
  In fact, our method computes the mean over all clean
  sub-reconstructions indicated by \(J\in \J\).

  The obtained results may be used to guide implementation in practice.
Equation~\eqref{eq:output-is-clean} explains how to choose subsets
\(\J\).
First of all, the mean of the clean sub-reconstructions
\(\nicefrac{1}{\abs{\J}}\sum_{J\in\J}\rcleanrv_J\) must resemble
the desired clean image.
This can be achieved by choosing \(\J\) to be a partition of \(\{1,
\ldots, m\}\), or, by choosing \(\J\) such that each measured data point
is contained in the same number of overlapping subsets
\(J\in\J\).
Not doing so introduces a systematic bias into the reconstruction.

Second, the sub-reconstructions should be homogeneously informative
throughout the image.
If the sub-reconstructions are very different, or contain limited
information about large parts of the image, then many dissimilar clean
images are consistent with the observed noisy reconstruction, and the
average over all these images will become blurred.

We note that \(\rcleanrv\) denotes the clean reconstruction, rather than
the unknown image.
This has two consequences.
  First, the theory predicts that artifacts that are unrelated to the
  measurement noise, e.g.\ under-sampling artifacts and reconstruction
  artifacts, will not be removed by the proposed network.
Second, if the reconstruction method also performs denoising
operations, for instance by blurring, then the result of our method
might become blurred.
  The same effect might occur when a non-linear reconstruction method
  is used, for which Proposition 1 does not generally hold.
  In this case, the regression function averages the bias introduced
  by the non-linear reconstruction of the noise.
In the next section, we use the considerations discussed above to
devise an approach for computed tomography.

\subsection{Noise2Inverse for computed tomography}\label{sec:noise2inverse-tomography}

\begin{figure}
  \centering
  \includegraphics[]{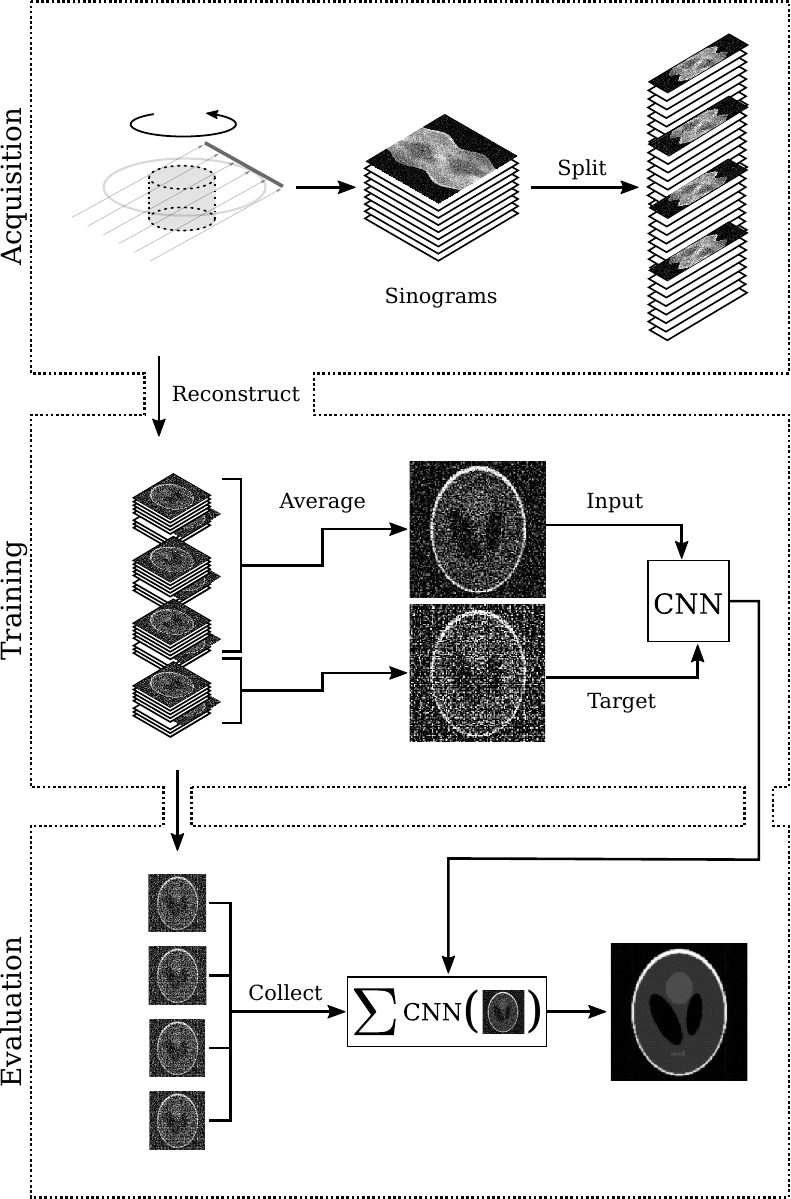}
  \caption{
    Noise2Inverse for computed tomography.
    First, 3D parallel-beam tomography obtains a stack of noisy sinograms
    by integrating over parallel lines at
    several angles.
    Next, the stack of sinograms is split along the angular axis.
    Then, the split sinograms are reconstructed to act as training
    dataset.
    During training, a dynamic subset of slices is averaged to form
    the input; the target is the average of the remaining slices.
    To obtain a low-noise result, the trained CNN is applied to all
    arrangements of input slices and averaged.
  }\label{fig:noise2inverse-option}
\end{figure}

In this section, we describe our implementation of Noise2Inverse for
3D parallel-beam tomography, and discuss how the implementation
relates to the theoretical considerations discussed before.

The 3D parallel-beam tomography problem may be considered as a stack
of 2D parallel-beam problems.
In 2D parallel-beam tomography, a parallel X-ray beam penetrates an
object, after which it is measured on a line detector.
The line detector rotates around the object while capturing the
intensity of the attenuated X-ray beam, as illustrated in the top
panel of Figure~\ref{fig:noise2inverse-option}.

In practice, a finite number of \(\numangles\) projections are
acquired on a line grid of \(\numpixelsy\) detector elements at fixed
angular intervals.
Hence, the projection data can be described by a vector
\(\pnoisy \in \Y = \R^m, m = \numangles \times \numpixelsy\), which is
known as the \emph{sinogram}.
Likewise, the two-dimensional imaged object is represented by a vector
\(\x \in \X = \R^n, n=\numpixelsx^2\).
We can formulate 2D parallel-beam tomography as a discrete linear
inverse problem, where \(\A=(a_{ij})\) is an \(m \times n\) matrix
such that \(a_{ij}\) represents the contribution of object pixel \(j\)
to detector pixel \(i\).
In 3D tomography, a sequence of 2D projection images of the 3D
structure is acquired, which may be converted to a stack of 2D
sinograms.

The imaged object can be recovered from the sinogram by a
reconstruction algorithm, such as the filtered back-projection
algorithm (FBP)~\cite{buzug-2008-comput-tomog}.
FBP is an example of a linear operator that couples the measured noise
in the reconstruction, as described in Equation~\eqref{eq:inverse-problems-reconstructions}.
In addition, it is typically fast to compute, although its
reconstructions tend to be noisy~\cite{chang-2013-asses-model}.

The Noise2Inverse method is well-suited to denoise this kind
of problem.
Suppose we have obtained a stack of 2D noisy sinograms
\(\pnoisyinstance_1, \pnoisyinstance_2, \ldots, \pnoisyinstance_N\),
acquired from a range of \(\numangles\) equally-spaced angles
\(\angle_1, \angle_2, \ldots, \angle_{\numangles}\).
Our approach follows the following steps.

First, we split each sinogram $\pnoisyinst_i$ into $\numsplits$ sub-sinograms
$\pnoisyinst_{i, 1}, \ldots, \pnoisyinst_{i, \numsplits}$ such that each
sub-sinogram $\pnoisyinst_{i,j}$ contains pixels from every $\numsplits$th
angle $\angle_j, \angle_{j+K}, \angle_{j+2K}, \ldots, \angle_{j +
  \numangles - \numsplits}$.
The number of splits \(\numsplits\) is a hyper-parameter of the method.

Using the FBP algorithm, we compute sub-reconstructions
\begin{align}
  \rnoisyinst_{i, j} = \Rec_j (\pnoisyinst_{i, j}), \quad j=1, \ldots, \numsplits.
\end{align}
For training, the division of the sub-reconstructions over the input and target is
determined by a collection $\J$, which contains subsets $J \subset \{1,
\ldots, \numsplits\}$.
For $J\subset \{1, \ldots, \numsplits\}$, we define the mean sub-reconstruction as
\begin{align}
  \rnoisyinst_{i, J} = \frac{1}{\abs{J}} \sum_{j \in J} \rnoisyinst_{i, j}.
\end{align}
As before, training of the neural network \(\network\) aims to find
\begin{align}
\label{eq:training-task-repeat-again}
\paramhat = \argmin_{\param}
\sum_{i=1}^N \sum_{J \in \J}
    \norm{\network\left(\rnoisyinst_{i, J^C}\right) - \rnoisyinst_{i, j}}_2^2.
\end{align}
  The final output, $\rcleaninst_{i, \text{out}}$, is computed slice
  by slice by section-wise averaging of the output of the trained network
  \begin{align*}
    \rcleaninst_{i, \text{out}} &= \frac{1}{\abs{\J}} \sum_{J \in \J} \trainednetwork\left(\rnoisyinst_{i, J^C}\right).
  \end{align*}
In this paper, we identify two training strategies specifying $\J$:
\begin{description}
\item[{X:1}]
  Using this strategy, the input is the mean of $\numsplits - 1$
  sub-reconstructions, and the target is the remaining
  sub-reconstruction, i.e.,
  \begin{align}
    \label{eq:x-1-strategy}
    \J_{X:1} = \{\{1\}, \{2\}, \ldots, \{\numsplits\}\}.
  \end{align}
\item[{1:X}]
  This is the reverse of the previous strategy: the input is a single
  sub-reconstruction, and the target is the mean of the remaining
  sub-reconstructions, i.e.,
  \begin{align}
    \label{eq:1-x-strategy}
    \J_{1:X} &= \{J^C  \mid J \in \J_{X:1} \}.
  \end{align}
\end{description}

In the 1:X strategy, the input is noisier than the target image, which
corresponds to supervised training, where the quality of the target
images is usually higher than the input images.
  The opposite is the case for the X:1 strategy, which corresponds more closely
  to Noise2Self denoising in its distribution of data between input and target,
  where more pixels are used to compute the input than to compute the target
  images.
  Note that other splits are possible, but we
  focus on these two strategies because they represent two extremes
  in the trade-off between input quality and target quality.

Our implementation of Noise2Inverse for tomography
is consistent with the theoretical considerations discussed in the
previous section.
In both strategies, we prevent biasing the reconstructions, by
ensuring that each projection angle occurs in reconstructions at the
same rate.
In fact, a property of FBP is that the full reconstruction is the mean
of the sub-reconstructions.
In theory, this means that training converges to the conditional
expectation of the \emph{full clean} FBP reconstruction.
Furthermore, we use every $\numsplits$th projection angle to compute the
reconstructions.
This ensures that the reconstructions are homogeneously informative
throughout the image, and we prevent missing wedge artifacts, which occur
when adjacent projection angles are used~\cite{midgley-2003-elect-micros}.
In addition, we use the FBP algorithm with the Ram-Lak
filter\cite{buzug-2008-comput-tomog}, which does not blur the
reconstructions to remove noise.
  Finally, we remark that our method is not geometry-specific, and can also be
  applied to non-parallel geometries, as is demonstrated in
  Section~\ref{sec:results-qualitative-comparison}.
In the next section, we describe the performance of this implementation in
practice.

\section{Results}\label{sec:results}

We performed several experiments on tomographic reconstruction
problems.
These experiments were performed with the aim of assessing the
performance of the proposed Noise2Inverse method, determining the
suitability of Noise2Self denoising for tomographic images, and
analyzing the impact of hyper-parameters on the performance of
Noise2Inverse.

  \textbf{Comparison to denoising techniques}
  Noise2Inverse is compared to tomographic
  reconstruction algorithms, an image denoising method, and an unsupervised deep
  learning method in Sections~\ref{sec:results-quantitative-comparison},
  \ref{sec:results-medical}, and~\ref{sec:results-qualitative-comparison}.
  These sections describe a quantitative evaluation on simulated tomographic data,
  medical CT data with simulated noise, and a qualitative evaluation on an
  existing experimental dataset.

\textbf{Noise2Self on tomographic images}
The experiments in Section~\ref{sec:results-noise2self} investigate a
transfer of Noise2Self denoising to inverse problems.
The Noise2Self method was evaluated on two datasets: one dataset with
noise common to tomographic reconstructions and one with similar but
element-wise independent noise.
In addition, Noise2Inverse was compared to several variations of
Noise2Self.

\textbf{Hyper-parameters}
In Section~\ref{sec:hyper-parameters}, the impact on the
reconstruction quality of several variables was investigated,
specifically, the number of projection angles $\numangles$, the number
of splits \(\numsplits\), the training strategy $\J$, and the neural
network architecture.
  In addition, we analyze the generalization performance of the Noise2Inverse
  approach by training on progressively smaller subsets of the training dataset.

We first describe the simulated tomographic dataset and our
implementation of Noise2Inverse.
Both are used throughout the experiments.

\textbf{Simulated data}
A cylindrical foam phantom was generated containing 100,000
randomly-placed non-overlapping bubbles.
Analytical projection images of the phantom were computed using the
open-source \texttt{foam\_ct\_phantom} package~\cite{pelt-2018-improv-tomog}.
The value of each detector pixel was calculated by taking the average
projection value of four equally-spaced rays through the pixel.
Projection images were acquired from $1024$ equally spaced angles.

The projection images of the foam dataset were corrupted with various
levels of Poisson noise.
The noise was varied by altering the average absorption of the sample
\(\absorption\) and the incident photon count per pixel \(I_0\).
The average absorption of the sample was calculated as the mean of the
vector \(1 - e^{-\pcleanvec_i}\) for positions \(i\) where
\(\pcleanvec_i\) was non-zero, and it was adjusted by modifying the
intensity of the sinogram.
The pixels in the noisy projections where sampled from
\(\rv{\tilde{p}}\), which for clean pixel value $\instance{p}$ was
distributed as
\begin{align*}
  I_0 e^{-\rv{\tilde{p}}} \sim \text{Poisson}\left(I_0 e^{-\instance{p}}\right).
\end{align*}
  i.e., a Poisson distribution on the pre-log raw data.
  Depending on the photon count and attenuation of the object,
  this type of noise is mean-zero conditional on the clean
  projections, as described in Equation (\ref{eq:noise-zero-mean}).

\begin{figure*}
  \centering
  \includegraphics[]{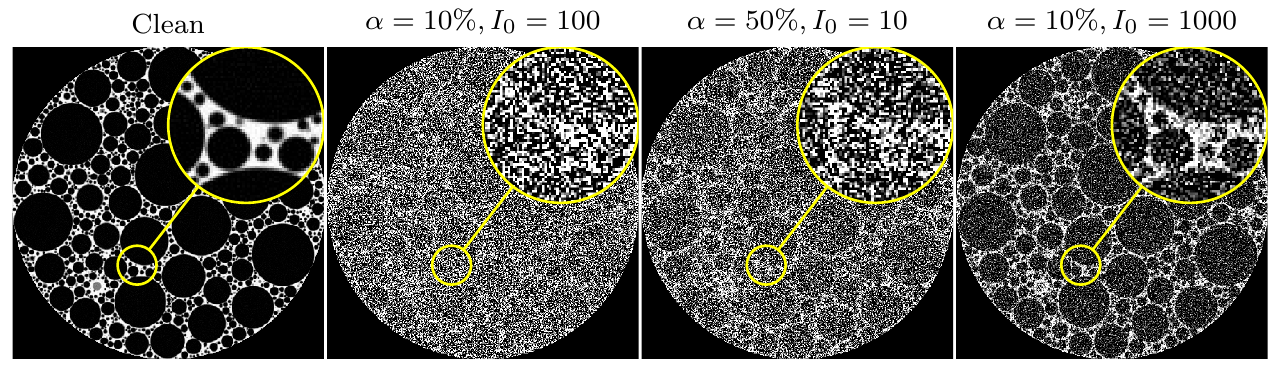}
  \caption{
    Displays of the clean reconstruction (left) and low-dose
    reconstructions of the central slice of the foam phantom.
    Both $\alpha$, the absorption of the phantom and $I_0$, the
    initial photon count per pixel, were varied.
    The yellow insets show an enlarged view of the reconstructions.
  }\label{fig:foam-comparison-noisy}
\end{figure*}

FBP reconstructions were computed on a \(512^3\) voxel grid with the
Ram-Lak filter using the ASTRA toolbox~\cite{aarle-2015-astra-toolb}.
On this grid, the radius of the random spheres ranged between \(1.5\)
and \(51\) voxels.
A reconstruction of the central slice of the foam phantom can be found
in Figure~\ref{fig:foam-comparison-noisy}, along with reconstructions
of the noisy projection datasets.

\textbf{Noise2Inverse}
We describe the Noise2Inverse implementation in terms of neural
network architecture and training procedure.

The principal network architecture used throughout the experiments was
the mixed-scale dense (MS-D) network~\cite{pelt-2017-mixed-scale}, of
which we used the open-source \texttt{msd\_pytorch} implementation~\cite{hendriksen-2019-msd-pytor}.
The MS-D network has 100 single-channel intermediate layers, and the
convolutions in layer \(i\) are dilated by \(d_i = 1 + (i \mod 10)\).
With 45,652 trainable network parameters, the MS-D architecture has
considerably fewer parameters than comparable network architectures,
reducing the risk of overfitting to the noise.
The MS-D architecture is compared with other architectures in Section~\ref{sec:hyper-parameters}.

The networks were trained for \(100\) epochs using the ADAM algorithm~\cite{kingma-2014-adam} with a mini-batch size of \(12\) and a learning
rate of \(10^{-3}\).

\subsection{Simulation study}\label{sec:results-quantitative-comparison}

In this section, Noise2Inverse is compared to two conventional iterative
reconstruction techniques: the simultaneous iterative reconstruction technique
(SIRT)~\cite{gregor-2008-comput-analy} and Total-Variation Minimization (TV-MIN)~\cite{beck-2009-fast-gradien}.
  In addition, we compare to the BM3D image denoising
  algorithm~\cite{dabov-2007-image-denois}, the Deep Image
  Prior\cite{ulyanov-2017-deep-image-prior}, and to supervised training.
The reconstruction quality of these methods is assessed on a simulated foam
phantom dataset with various noise profiles.

For Noise2Inverse, we used the X:1 training strategy with
\(\numsplits=4\) splits.
We show that this is a robust choice in Section~\ref{sec:hyper-parameters}.

\begin{figure*}
  \centering
  \includegraphics[]{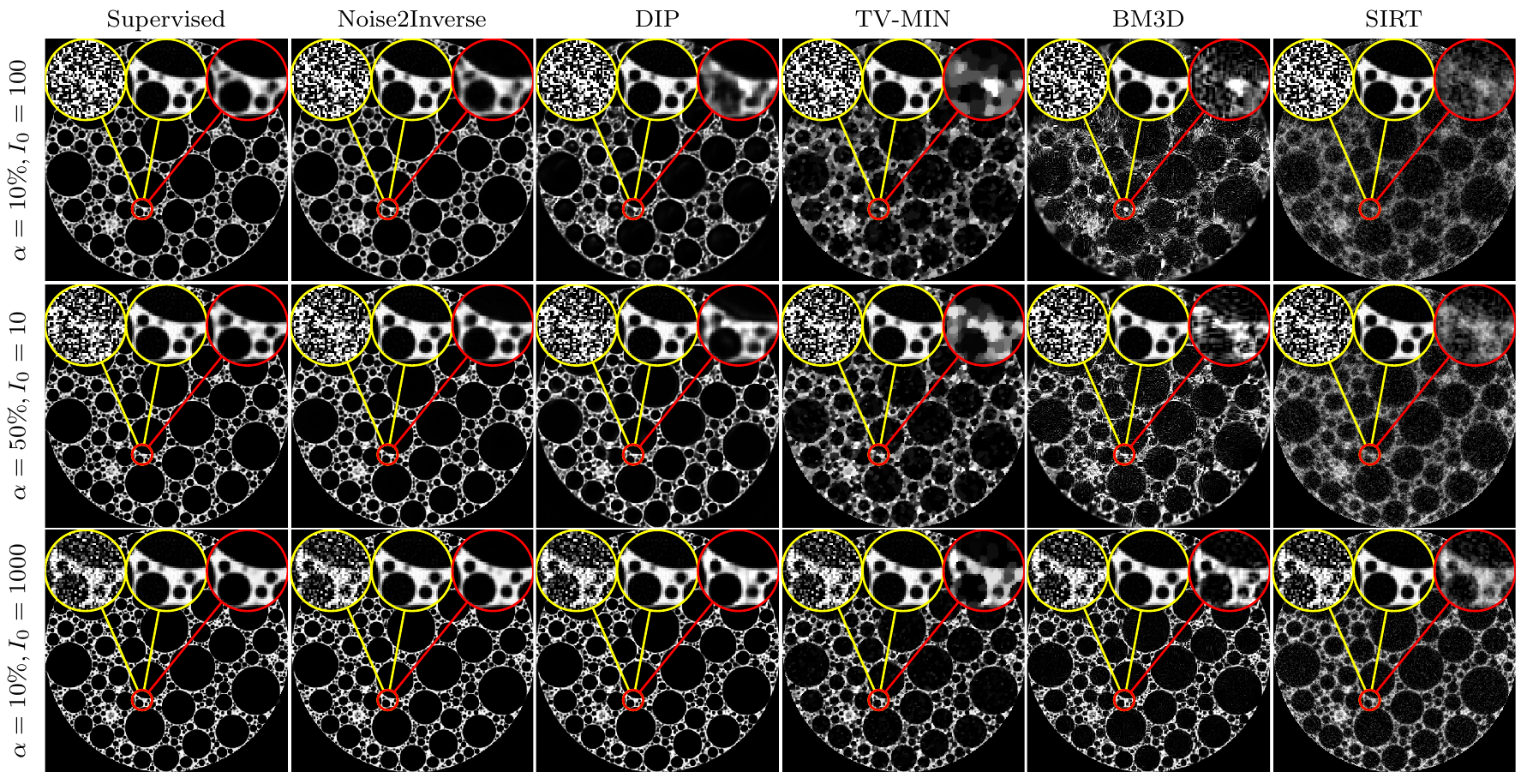}
  \caption{
    Results of supervised training, Noise2Inverse, Deep Image Prior (DIP),
    TV-MIN, BM3D, and SIRT on simulated foam phantoms with varying absorption
    $\absorption$ and photon count $I_0$.
    Results are shown on the central slice.
    The insets display the noisy and clean reconstructions (yellow)
    and the algorithm output (red).
  }\label{fig:foam-comparison-sirt-fista}
\end{figure*}

\textbf{Iterative reconstruction}
The hyper-parameters of SIRT and TV-MIN were tuned using the usually
unavailable clean reconstructions.
Therefore, the results of SIRT and TV-MIN might be better than what is
achievable in practice, but they serve as a useful reference for
comparison to Noise2Inverse.
SIRT has no explicit hyper-parameters, but its iterative nature can be
exploited for regularization: early stopping of the algorithm can
attenuate high-frequency noise in the reconstructed image~\cite{gregor-2008-comput-analy}.
We selected the number of iterations (with a maximum of 1000)
with the lowest Peak Signal to Noise Ratio (PSNR) on the central slice
with respect to the clean reconstruction.

The FISTA algorithm~\cite{beck-2009-fast-gradien} was used to
calculate the TV-MIN reconstruction.
TV-MIN has a regularization parameter $\lambda$ that effectively
penalizes steps in the gray value of the reconstructed image.
As with SIRT, we selected the optimal number of iterations (with a
maximum of 500) based on the PSNR of the central slice with respect to
the clean reconstruction, and the value of the \(\lambda\)
parameter
  maximizing the PSNR
was determined using the Nelder-Mead method~\cite{virtanen-2019-scipy}.

  \textbf{BM3D}
  We used the BM3D implementation described in~\cite{makinen-2019-exact-trans}.
  The BM3D algorithm was applied to the noisy FBP reconstructions and
  provided with the standard deviation of the noise, which was
  calculated from the difference image between the noisy and clean FBP
  reconstruction.
  The addition of a \emph{prewhitening} step can improve denoising
  performance~\cite{seghouane-2014-prewh-high}, but was not included as its
  computation becomes infeasible for large image sizes.

  \textbf{Supervised}
  A separate training dataset was created to train MS-D networks with
  a supervised training approach.
  Here, the input and target images were noisy and clean
  reconstructions, respectively.
  The training parameters for supervised training --- learning rate,
  batch size, network architecture --- were exactly the same as for
  the Noise2Inverse network.

  \textbf{Deep Image Prior}
  We used the Deep Image Prior implementation from~\cite{ulyanov-2017-deep-image-prior}.
  The quality of the result can be improved by adding noise to
  the input and by employing an exponentially decaying average of
  recent iterations~\cite{cheng-2019-bayes-persp}.
  We used both techniques.
  To maximize the PSNR with respect to the ground truth, the training
  is stopped early with a maximum of $10000$ iterations, and the
  $\sigma$ parameter of the input noise is optimized using a line
  search.

\begin{table}[]
  \centering
  \caption[]{
    On the full volume and on the central slice:
    comparison of PSNR and SSIM metrics for SIRT, TV-MIN, BM3D,
    Deep Image Prior, Noise2Inverse, and a supervised CNN at several
    noise profiles.
    Bold font is used to emphasize the best metrics, excluding
      supervised training, which serves as an oracle case for comparison.
  }\label{tab:foam-comparison}
\begin{tabular}{rrlrrrr}
\toprule
                      &                       & {}               & \multicolumn{2}{c}{Full Volume} & \multicolumn{2}{c}{Central slice} \\
                      &                       &                  & PSNR                            & SSIM     & PSNR      & SSIM       \\
$\absorption$         & $I_0$                 & Method           &                                 &          &           &            \\
\midrule
\multirow{6}{*}{10\%} & \multirow{6}{*}{100}  & Supervised       & 20.01                           & 0.83     & 20.02     & 0.80       \\
                      &                       & Noise2Inverse    & \bf 19.71                       & \bf 0.78 & \bf 19.63 & \bf 0.74   \\
                      &                       & Deep Image Prior &                                 &          & 17.98     & 0.59       \\
                      &                       & TV-MIN           & 16.89                           & 0.46     & 16.78     & 0.40       \\
                      &                       & BM3D             & 14.79                           & 0.38     & 14.81     & 0.33       \\
                      &                       & SIRT             & 15.56                           & 0.36     & 15.54     & 0.32       \\
\midrule
\multirow{6}{*}{50\%} & \multirow{6}{*}{10}   & Supervised       & 21.77                           & 0.86     & 21.71     & 0.83       \\
                      &                       & Noise2Inverse    & \bf 21.66                       & \bf 0.79 & \bf 21.62 & \bf 0.75   \\
                      &                       & Deep Image Prior &                                 &          & 19.75     & 0.67       \\
                      &                       & TV-MIN           & 18.08                           & 0.53     & 17.99     & 0.48       \\
                      &                       & BM3D             & 16.65                           & 0.49     & 16.74     & 0.45       \\
                      &                       & SIRT             & 16.53                           & 0.42     & 16.50     & 0.37       \\
\midrule
\multirow{6}{*}{10\%} & \multirow{6}{*}{1000} & Supervised       & 26.55                           & 0.91     & 26.50     & 0.88       \\
                      &                       & Noise2Inverse    & \bf 26.25                       & \bf 0.89 & \bf 26.24 & \bf 0.87   \\
                      &                       & Deep Image Prior &                                 &          & 24.03     & 0.86       \\
                      &                       & TV-MIN           & 21.24                           & 0.68     & 21.24     & 0.61       \\
                      &                       & BM3D             & 21.14                           & 0.69     & 21.11     & 0.65       \\
                      &                       & SIRT             & 18.84                           & 0.53     & 18.82     & 0.48       \\
\bottomrule
\end{tabular}
\end{table}

\textbf{Metrics and evaluation}
The output of each method was compared to the clean FBP reconstruction
using two metrics: the structural similarity index (SSIM)~\cite{wang-2004-image-qualit-asses} and the Peak Signal to Noise Ratio
(PSNR).
Because the reconstructed images did not fall in the \([0,1]\) range,
these metrics were computed with a data range that was determined by
the minimum and maximum intensity of the clean reconstructed images.
  The metrics were calculated on the convex hull surrounding the object, which
  diminishes the importance of the background image quality.
Due to the computational demands of deep image prior, we compute
metrics on a single slice of the reconstruction rather than on the
whole volume.

The top row of Figure~\ref{fig:foam-comparison-sirt-fista} displays
the output of Noise2Inverse for the central slice of the three simulated
datasets.
Denoising these datasets is challenging, as can be seen when comparing
with SIRT and TV-MIN:\ these algorithms fail to recover several fine
details.
In contrast, our method achieves a much improved visual impression on
all three datasets.
As can be seen in Table~\ref{tab:foam-comparison}, the PSNR and SSIM
metrics of the Noise2Inverse method are considerably higher.
  The supervised network attains the best metrics, although
  by a slight margin compared to the Noise2Inverse method.

\subsection{Medical CT}\label{sec:results-medical}

\begin{figure*}
    \centering
    \includegraphics[]{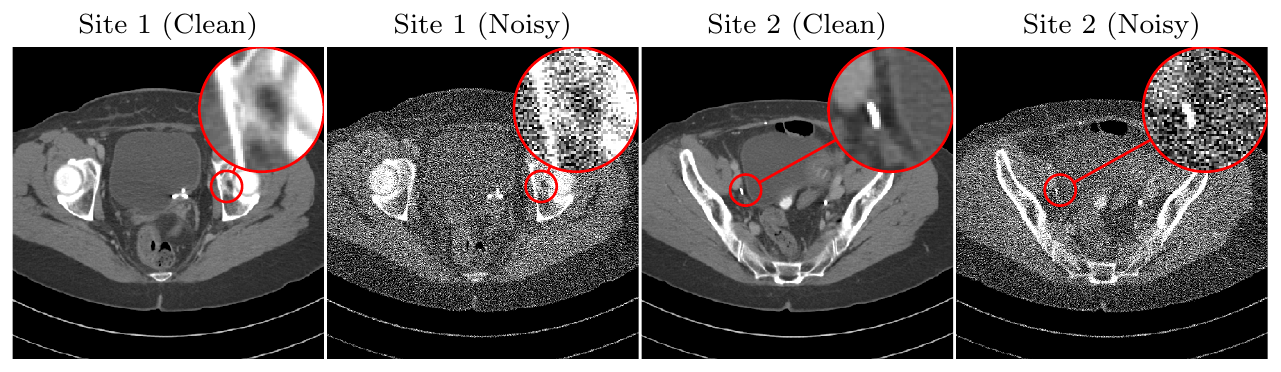}
    \caption{Original high-dose reconstructions of low-dose CT grand challenge
      (clean) and reconstructions with simulated noise (noisy). }\label{fig:mayo-data}
  \end{figure*}

  To assess the quality of reconstruction on medical data, we evaluate our
  method on simulated data from human abdomen CT scans from the low-dose CT
  Grand Challenge dataset~\cite{mccollough-2017-low-dose}.
  This dataset contains full-dose reconstructions of 10 patients, consisting of
  a total of $2378$ slices of $512 \times 512$ pixels.
  Following~\cite{adler-2018-learn-primal}, sinograms were computed from these
  reconstructions by projecting onto a fan-beam geometry.
  Noise was applied, corresponding to a photon count of $10,000$ incident
  photons per pixel.
  Reconstructions are shown in Figure~\ref{fig:mayo-data}.

\begin{figure*}[]
  \centering
  \includegraphics[]{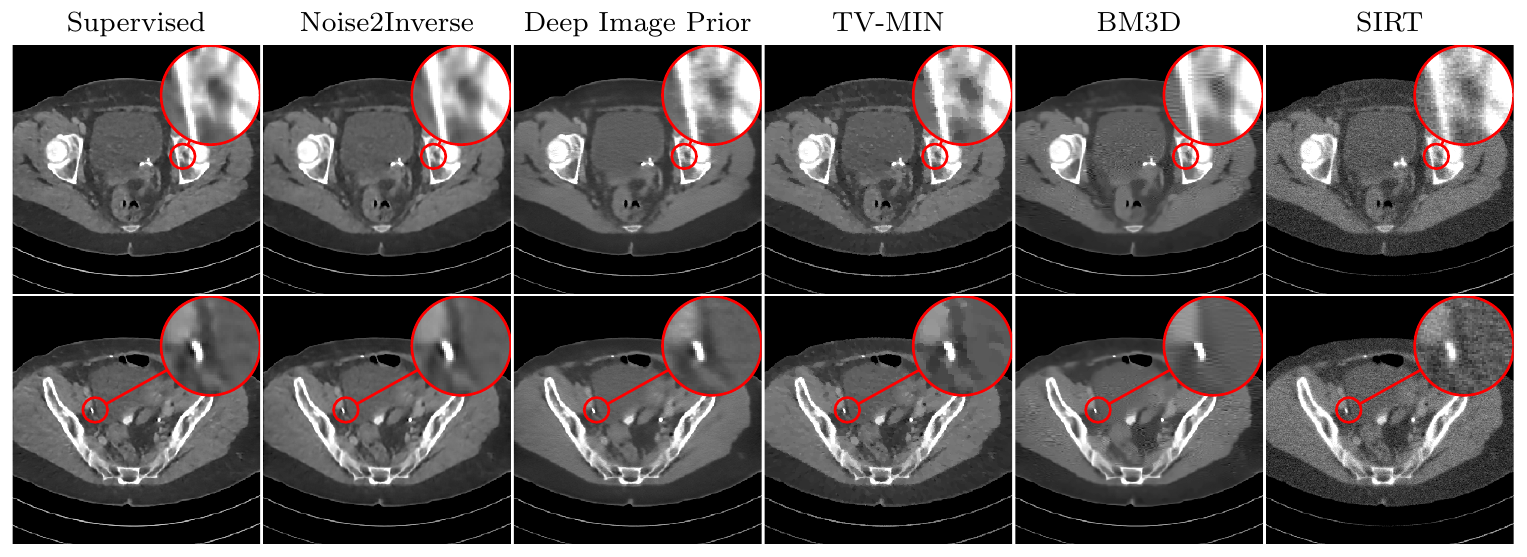}
  \caption{
    Results of supervised training, Noise2Inverse, Deep Image Prior, TV-MIN, BM3D, and
    SIRT on Low-dose CT grand challenge data with simulated noise.
    The red insets display the algorithm output.
  }\label{fig:mayo-comparison}
\end{figure*}

  We compare the same methods as before. The dataset was split into a training
  dataset, consisting of nine patients, and a test set, containing the remaining
  patient. Both Noise2Inverse and the supervised CNN were trained on the
  training set. The optimal hyperparameters for SIRT, TV-MIN, and BM3D were
  determined on the training set. The Deep Image Prior, including its
  hyperparameters, was directly optimized with respect to the slices displayed
  in Figure~\ref{fig:mayo-comparison}. Metrics were calculated on the full
  volume of the test patient, and on the top displayed slice in
  Figure~\ref{fig:mayo-comparison}.

  Results are shown in Figure~\ref{fig:mayo-comparison} and
  Table~\ref{tab:mayo-comparison}.
  The Noise2Inverse method achieves similar results to TV-MIN, but without the
  staircasing artifacts.
  The difference between the methods is smaller in this experiment.
  For the SSIM metric, this is likely due to the low contrast of structures of
  interest compared to the full intensity range of the
  reconstructions.
  In general, compared to previous experiments, the noise has significantly lower
  intensity, and many different objects structures are present, each of
  which must be learned by the neural network.

\begin{table}[]
  \centering
  \caption[]{
    Medical data:
    comparison of PSNR and SSIM metrics for SIRT, TV-MIN, BM3D, Deep Image
    Prior, a supervised CNN, and Noise2Inverse.
    Bold font is used to emphasize the best metrics, excluding
      supervised training, which serves as an oracle case for comparison.
  }\label{tab:mayo-comparison}

  \begin{tabular}{lrrrr}
    \toprule
    {}               & \multicolumn{2}{c}{Full volume} & \multicolumn{2}{c}{Single slice} \\
                     & PSNR                            & SSIM     & PSNR      & SSIM      \\
    Method           &                                 &          &           &           \\
    \midrule
    Supervised       & 46.34                           & 0.99     & 46.29     & 0.99      \\
    Noise2Inverse    & \bf 45.06                       & \bf 0.99 & 45.46     & \bf 0.99  \\
    TV-MIN           & 44.91                           & \bf 0.99 & \bf 45.65 & 0.98      \\
    Deep Image Prior &                                 &          & 44.57     & 0.98      \\
    BM3D             & 43.84                           & \bf 0.99 & 43.97     & 0.98      \\
    SIRT             & 39.87                           & 0.97     & 40.61     & 0.95      \\
    \bottomrule
  \end{tabular}

\end{table}

\subsection{Experimental data}\label{sec:results-qualitative-comparison}

The Noise2Inverse method was compared to SIRT and TV-MIN on an
existing real-world experimental dataset from TomoBank~\cite{de-2018-tomob}.
The dataset, \texttt{Dorthe\_F\_002}, was acquired at the Advanced
Photon Source at Argonne National Laboratory, and contained \(900\)
noisy projection images of \(960 \times 600\) pixels depicting a
cylinder of glass beads that was scanned at experimental conditions
designed to capture the dynamics of fast evolving samples.
At $6$ milliseconds per projection image, the exposure time was
therefore much shorter than what is required for low-noise data
acquisition~\cite{de-2018-tomob}.
The data was pre-processed with the TomoPy software package~\cite{guersoy-2014-tomop} and reconstructed with
FBP\cite{aarle-2015-astra-toolb}, resulting in \(900\) 2D slices of
\(960 \times 960\) pixels.
We stress that no low-noise projection images were available.

For Noise2Inverse, an MS-D network was trained with the X:1 strategy
and \(4\) splits for \(100\) epochs.
The best parameter settings for SIRT and TV-MIN were determined by
visual inspection.
For SIRT, the best reconstruction was chosen from \(1000\) iterations on
the central slice.
For TV-MIN, the number of iterations was fixed at \(500\), and the optimal
value of the regularization parameter was chosen from
  several values regularly spaced on an exponential grid.
For BM3D, the best image was chosen from various values of the
standard deviation parameter.
  We have omitted the Deep Image Prior since there was no ground truth with
  respect to which to perform early stopping.

\begin{figure*}[]
  \centering
  \includegraphics[]{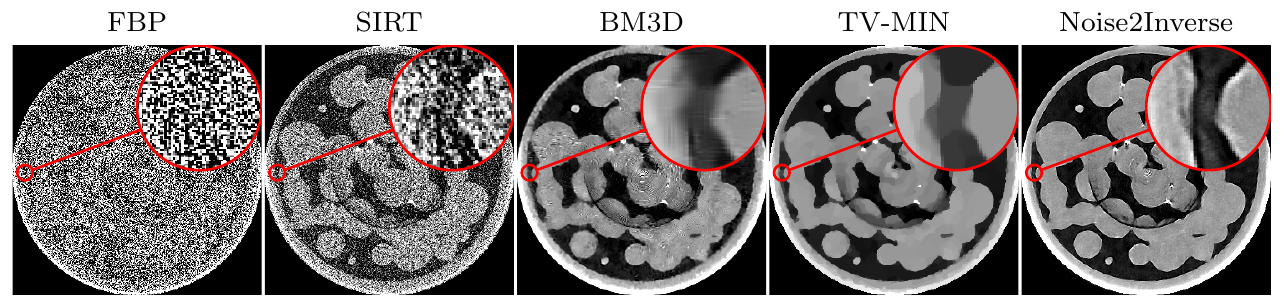}
  \caption{
    Reconstructions of cylinder containing glass beads~\cite{de-2018-tomob} using: FBP, SIRT, BM3D, TV-MIN, and the proposed
    Noise2Inverse method.
    The red insets show an enlarged view of the algorithm output.
  }\label{fig:tomobank-comparison}
\end{figure*}

After initial reconstructions, we found that the reported value of the
center of rotation offset --- \(4.5\) pixels from center --- yielded
unsatisfactory results.
The reconstructions in Figure~\ref{fig:tomobank-comparison} were
computed with a center of rotation that was shifted by \(8.9\) pixels.
Results are shown for the central slice of the reconstructed volume.
The FBP and SIRT reconstructions exhibit severe noise.
The TV-MIN reconstruction improves on the level of noise, but contains
stepping artifacts that reduce the effective resolution.
Our method is able to remove the noise while retaining the finer
structure of the image.

\subsection{Self-supervised image denoising for tomography}\label{sec:results-noise2self}

The performance of Noise2Self on tomographic images was evaluated in
two experiments.
The first experiment tested the element-wise independence requirement,
by evaluating Noise2Self on images corrupted by element-wise independent
noise and on images reconstructed from noisy projection data.
The second experiment was a comparison of Noise2Inverse to Noise2Self,
including variations of Noise2Self applied to projection and sinogram
images.
We first describe the Noise2Self implementation.

\textbf{Noise2Self}
The original implementation of Noise2Self~\cite{batson-2019-noise2}
was used, which obtains better performance than the simplified scheme
discussed in Section~\ref{sec:deep-learning}.
The training procedure was the same as for Noise2Inverse: an MS-D
network was trained for 100 epochs as described at the beginning of
Section~\ref{sec:results}.

\begin{figure*}[]
  \centering
  \includegraphics[]{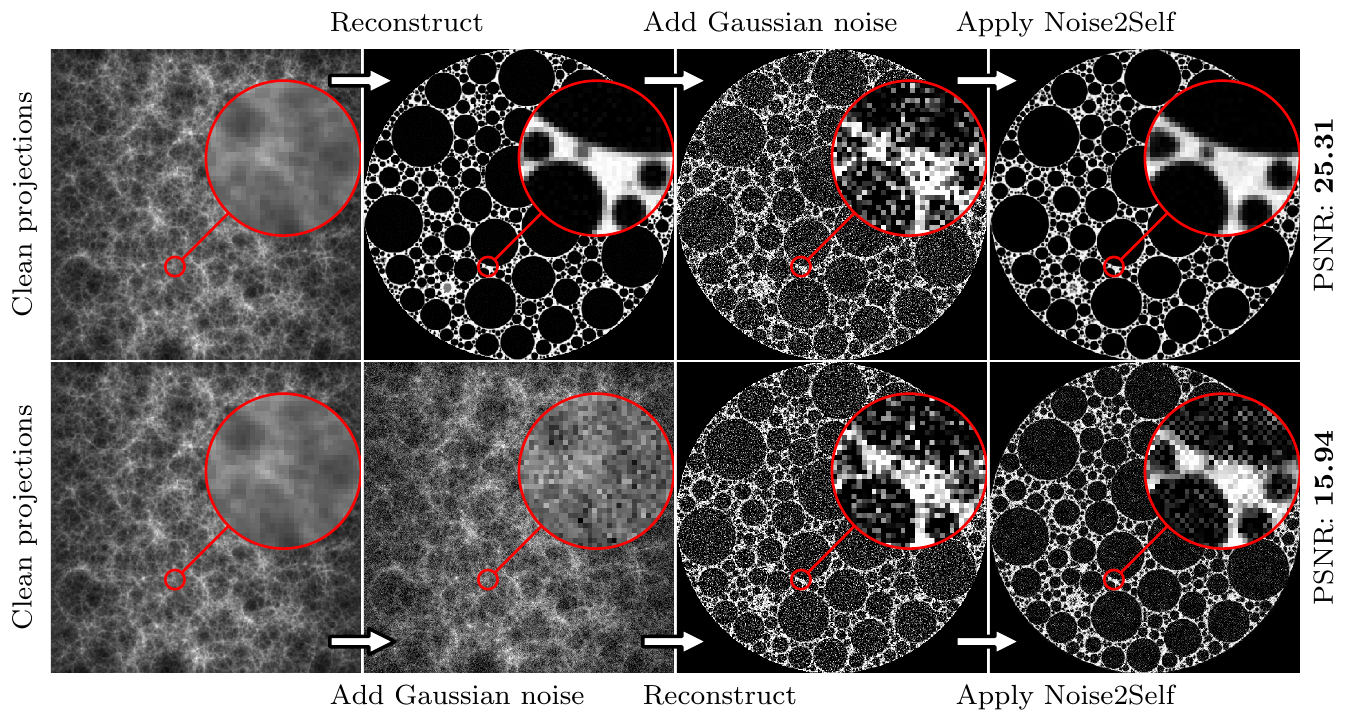}
  \caption{
    The effect of element-wise independence of the noise on the
    Noise2Self method.
    In the top row, Gaussian noise is added to a reconstruction, and
    Noise2Self is applied to remove it.
    In the bottom row, Gaussian noise is added to the projections
    before reconstruction, resulting in a reconstructed image with
    similar but coupled noise.
    Noise2Self achieves lower PSNR in the bottom row than in the top
    row.
  }\label{fig:noise2self}
\end{figure*}

\textbf{Tomographic versus photographic noise}
Noise2Self was applied to images with coupled reconstructed noise and to
similar but element-wise independent noise.
In these experiments, the same foam phantom was used as before, and
Gaussian noise was used throughout the comparison to strictly compare
the independence properties of the noise.
First, we confirmed that Noise2Self obtained denoised images when the
noise satisfied the element-wise independence property.
In this first case, a clean reconstruction was computed on a \(512^3\)
voxel grid, and independent and identically distributed (i.i.d.)
Gaussian noise was added to the reconstructed images.
The PSNR of the noisy volume with respect to the clean reconstruction
was \(11.06\).
Then, Noise2Self was applied to obtain a denoised volume with
significantly improved PSNR of \(25.23\).
This process is displayed in the top row of
Figure~\ref{fig:noise2self}.

Next, we investigated how Noise2Self performed on coupled
reconstructed noise.
In this case, i.i.d.\ Gaussian noise was added to the projection
images, and a reconstruction was computed afterwards.
The PSNR of this noisy reconstruction with respect to the clean
reconstruction was \(11.59\).
When Noise2Self was applied to the noisy reconstructed volume, it
obtained a PSNR of \(16.14\), which is only half of the improvement
that it obtained in the first case.
This process is displayed in the bottom row of Figure~\ref{fig:noise2self}.

The results displayed in Figure~\ref{fig:noise2self} demonstrate that
the performance of Noise2Self is substantially degraded when the noise
is not element-wise independent.
Even though the starting PSNR in the bottom row is slightly higher,
the PSNR improvement is only half of the top row.
In the top row, the validation error continued to improve for 100
epochs, whereas in the bottom row, training started to overfit to the noise
within the first 10 epochs of training, which could be caused by the
statistical dependence between the input and target images.

\textbf{Noise2Self on sinogram and projections}
To mitigate the effect of coupled noise, Noise2Self was also applied
to images that do satisfy the pixel-wise independence property: the
projection images and sinograms.
In these cases, Noise2Self was first applied to denoise the raw
images, and reconstructions were computed from the denoised projection
images or sinograms.

\begin{figure}[]
  \centering
  \includegraphics[width=\columnwidth]{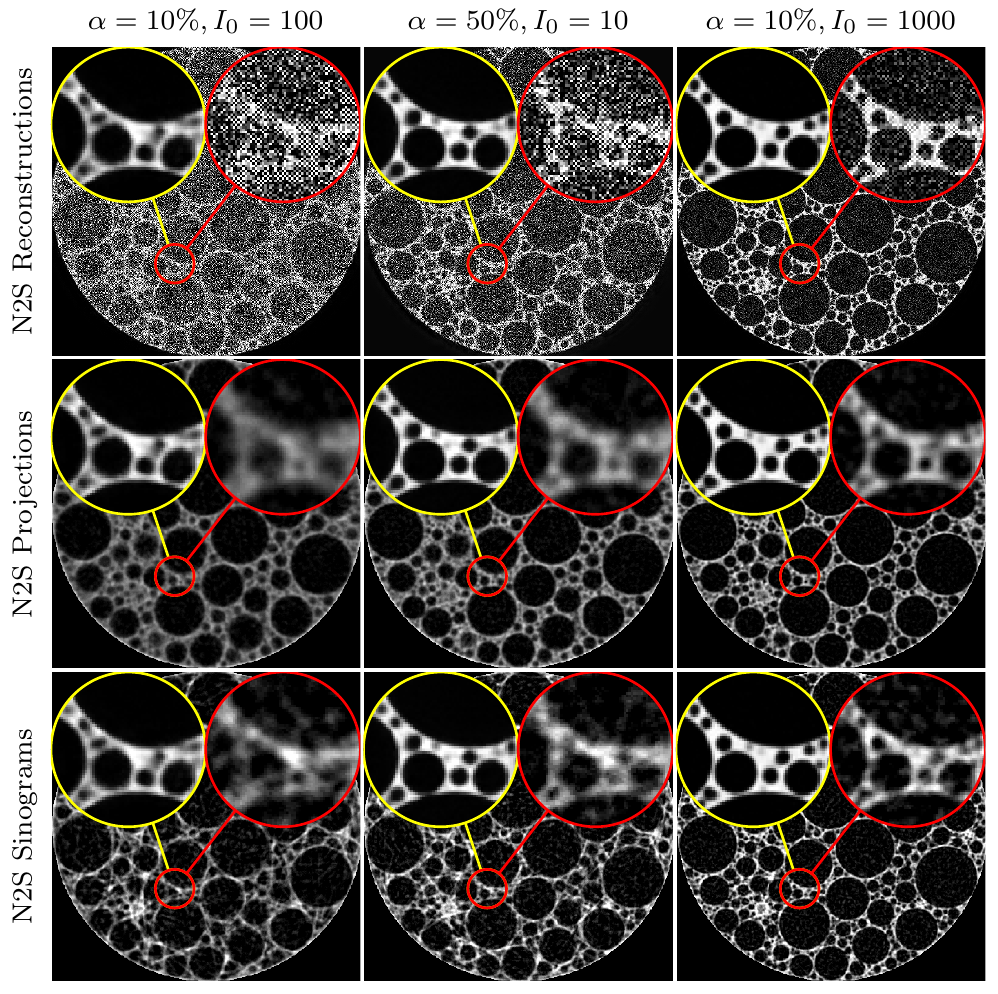}
  \caption{
    From top to bottom, results on the central slice of the foam
    phantom of Noise2Self applied to reconstructed, projection, and
    sinogram images.
    For comparison, the insets show the output of Noise2Inverse
    (yellow) and Noise2Self (red).
  }\label{fig:foam-comparison-denoising}
\end{figure}

As can be seen in Figure~\ref{fig:foam-comparison-denoising}, the
variations of Noise2Self did improve results, but not beyond
Noise2Inverse.
Although applying Noise2Self on the projection and sinogram images did
accurately denoise the raw images, the resulting reconstructions of
these denoised images exhibited some blurring (projections) and
streaks (sinograms).
% DISCUSS
% that could result from the fact that the consistency of the projection
% images with respect to the forward operator is not necessarily
% preserved.
% DISCUSS
% Similar results were observed by others \cite{buchholz-2019-cryo-care}.
As displayed in Table~\ref{tab:denoising-comparison}, the
Noise2Self-based method with the best metrics, Noise2Self on
sinograms, obtains PSNR on par with TV-MIN and SSIM worse than TV-MIN,
see Table~\ref{tab:foam-comparison}.

\begin{table}[]
  \centering
  \caption{
    Comparison of PSNR and SSIM metrics for Noise2Self on reconstruction,
    projection, and sinogram images.
  }\label{tab:denoising-comparison}
  \begin{tabular}{rrlrr}
\toprule
    Absorption        & $I_0$                 & Method              & PSNR      & SSIM     \\
\midrule
\multirow{3}{*}{10\%} & \multirow{3}{*}{100}  & N2S Reconstructions & 6.37      & 0.27     \\
                      &                       & N2S Projections     & 16.43     & 0.44     \\
                      &                       & N2S Sinograms       & \it 16.98 & \it 0.45 \\
                      &                       & Noise2Inverse       & \bf 19.71 & \bf 0.78 \\
\midrule
\multirow{3}{*}{50\%} & \multirow{3}{*}{10}   & N2S Reconstructions & 9.12      & 0.20     \\
                      &                       & N2S Projections     & 17.49     & 0.49     \\
                      &                       & N2S Sinograms       & \it 18.06 & \it 0.51 \\
                      &                       & Noise2Inverse       & \bf 21.66 & \bf 0.79 \\
\midrule
\multirow{3}{*}{10\%} & \multirow{3}{*}{1000} & N2S Reconstructions & 15.39     & 0.50     \\
                      &                       & N2S Projections     & 19.57     & \it 0.62 \\
                      &                       & N2S Sinograms       & \it 20.62 & 0.60     \\
                      &                       & Noise2Inverse       & \bf 26.25 & \bf 0.89 \\
\bottomrule
\end{tabular}
\end{table}

\subsection{Noise2Inverse and missing wedge artifacts}\label{sec:missing-wedge}

  The quality of tomographic reconstructions may be degraded due to artifacts
  other than measurements noise, such as missing wedge artifacts.
  These artifacts arise when projection data is acquired along an arc spanning
  less than 180\textdegree.
  The theoretical results in Section~\ref{sec:convergence} predict that
  Noise2Inverse preserves these artifacts.

\begin{figure}[]
  \centering
  \includegraphics[width=\columnwidth]{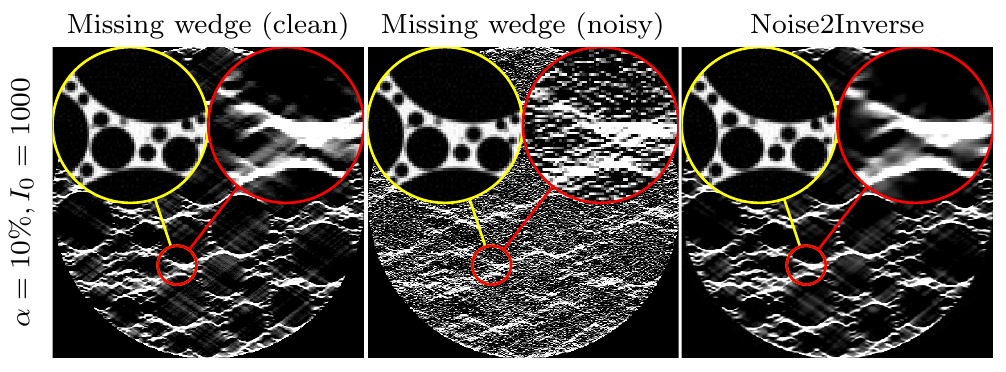}
  \caption{
    Noise2Inverse applied to a noisy dataset with missing wedge artifacts.
    The red and yellow insets show an enlarged view of the output and
    ground-truth, respectively.
  }\label{fig:foam-missing-wedge}
\end{figure}

  To test this prediction, we apply Noise2Inverse to a foam dataset where the
  reconstructions are computed from 400 projection images along an arc of
  approximately 60\textdegree.
  Noise is applied consistent with an absorption of $10\%$ and an incident photon
  count of $1000$ photons per pixel.
  As can be seen in Figure~\ref{fig:foam-missing-wedge}, Noise2Inverse
  accurately denoises the reconstructed image, but leaves the missing wedge
  artifacts intact.

\subsection{Hyper-parameters}\label{sec:hyper-parameters}

We analyzed the influence of the number of splits, training strategy,
number of projection angles, and neural network architecture on the
effectiveness of Noise2Inverse.
In addition, we tested the generalization by training on subsets of
the data.

The same foam phantom was used, and noisy projection data were acquired
from \(512\), \(1024\), and \(2048\) angles, of which the first and
last acquisitions were under-sampling and over-sampling the projection
angles, respectively.
For each dataset, the total number of incident photons remained
constant: we used \mbox{$I_0=400, 200, 100$} for
\(\numangles=512, 1024, 2048\), respectively.
The average absorption was \(23\%\), which is the default value of the
\texttt{foam\_ct\_phantom} package.

\textbf{Splits and strategy}
The performance of the Noise2Inverse method was evaluated
with a number of splits \(K = 2,4,8,16, 32\), and with
strategies X:1 and 1:X, see Equations~\eqref{eq:x-1-strategy} and~\eqref{eq:1-x-strategy}.
These experiments were performed with MS-D networks, which were
trained for 100 epochs, and used the same training procedure as
before.

\begin{figure}[]
  \centering
  \includegraphics[]{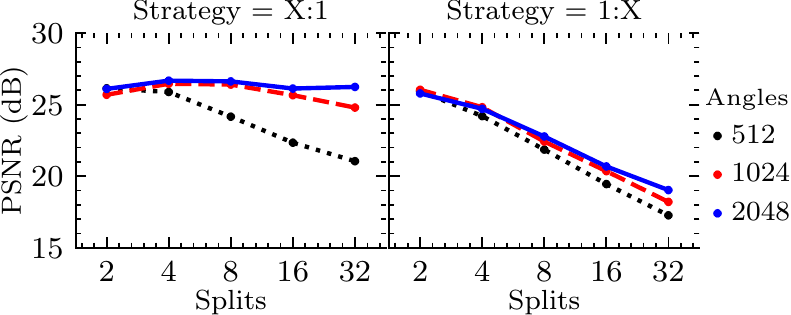}
  \caption{
    The PSNR metric for the Noise2Inverse method with the MS-D network
    applied on the foam phantom with varying number of splits, angles,
    and varying input-target splitting strategies.
    The X:1 strategy attains higher PSNR than the 1:X strategy.
  }\label{fig:x2y-comparison-splits-angles}
\end{figure}

The PSNR metrics are displayed in Figure~\ref{fig:x2y-comparison-splits-angles}.
The figure shows that the X:1 strategy yields considerably better
results than the 1:X strategy, except for \(K=2\), where they are
equivalent.
Setting the number of splits to \(\numsplits=2\) yields good results
across the board, but the PSNR can be improved by setting \(\numsplits\)
to \(4\) or \(8\), if the projection angles are not
under-sampled.
In general, the figure shows that increasing the number of acquired
projection images can improve reconstruction quality without
increasing the photon count.
 On the other hand, we note that reducing the number
  of projection images further can reduce the reconstruction quality
  as the artifacts arising from undersampling are not
  removed by the neural network.

\textbf{Neural network architectures}
We compared three neural network architectures: the U-Net~\cite{ronneberger-2015-u-net}, DnCNN~\cite{zhang-2017-beyon-gauss-denois}, and the previously described
MS-D~\cite{pelt-2017-mixed-scale} network architectures, all of which
were implemented in PyTorch~\cite{paszke-2017-autom-pytor}.

The U-net is based on a widely available open source
implementation\footnote{\url{https://github.com/milesial/Pytorch-UNet/}},
which is a mix of the architectures described in~\cite{cicek-2016-u-net,ronneberger-2015-u-net}.
Like~\cite{ronneberger-2015-u-net}, the images are down-sampled four
times using \(2\times 2\) max-pooling, the ``up-convolutions'' have
trainable parameters, and the convolutions have \(3\times 3\) kernels.
Like~\cite{cicek-2016-u-net}, this implementation uses batch
normalization before each ReLU, the smallest image layers are 512
channels instead of 1024 channels, and zero-padding is used instead
of reflection-padding.
The resulting network has 14,787,777 trainable network parameters.

We used the DnCNN implementation from~\cite{batson-2019-noise2} with a
depth of 20 layers, which is advised for non-Gaussian denoising~\cite{zhang-2017-beyon-gauss-denois}.
The resulting network has 667,008 trainable network parameters.

\begin{figure}[]
  \centering
  \includegraphics[]{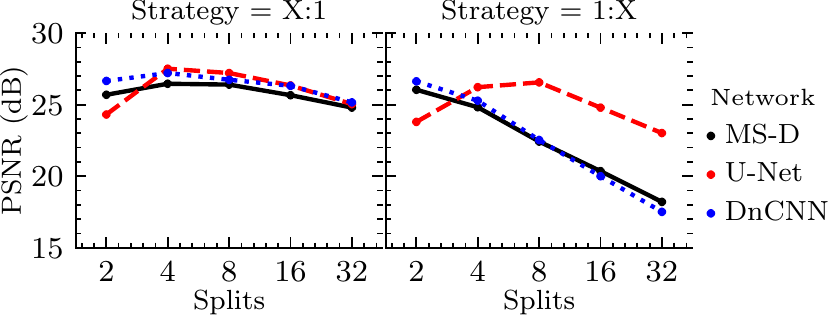}
  \caption{
    Comparison of the PSNR metric.
    The MS-D, U-Net, and DnCNN networks were trained for 100 epochs on
    the foam phantom with 1024 projection angles.
  }\label{fig:network-comparison}
\end{figure}

The previous experiment was repeated on the dataset containing 1024
projection images.
The networks were trained for 100 epochs, and used the same training
procedure as before.
The results are displayed in Figure~\ref{fig:network-comparison}.
The figure shows that the U-net achieved overall highest performance
using the X:1 strategy with \(4\) splits.
In addition, the effect of the number of splits \(\numsplits\) is roughly the same
across strategies and network architectures, except for U-net.
In fact, the PSNR metric of the U-Net with the 1:X strategy initially
increases when \(\numsplits\) is increased, which might be due to the
large network architecture and number of parameters compared to the
other two neural network architectures.
Nonetheless, the X:1 strategy consistently attains higher PSNR than
the 1:X for the U-net as well.
We note that the U-Nets performed worse than the other networks with 2
splits, which suggests that training might have overfit the noise.

\begin{figure}[]
  \centering
  \includegraphics[]{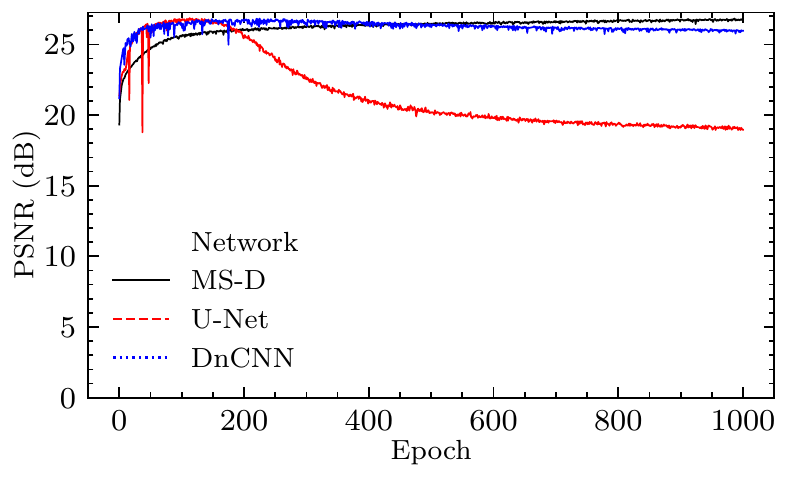}
  \caption{
    The PSNR on the central slice as training progressed.
    A U-Net, DnCNN, and MS-D network were trained with the X:1
    strategy and number of splits $\numsplits=4$ for 1000 epochs on
    the foam phantom reconstructed from 1024 projection angles.
  }\label{fig:network-overfit}
\end{figure}

\textbf{Overfitting}
We tested if the networks overfit the noise when trained for a long
time.
All three networks were trained for \(1000\) epochs using the X:1
strategy and \(\numsplits=4\) on the same foam dataset with 1024
projection angles.
The resulting PSNR on the central slice as training progressed is
displayed in Figure~\ref{fig:network-overfit}.
The figure shows that U-Net and DnCNN started to fit the noise,
whereas the PSNR of the MS-D network continued to increase.
  This matches earlier results on
  overfitting~\cite{pelt-2018-improv-tomog,pelt-2017-mixed-scale,hendriksen-2019-fly-machin}.
  If the training dataset had been larger, these effects could have been less
  pronounced.

\begin{figure}[]
  \centering
  \includegraphics[]{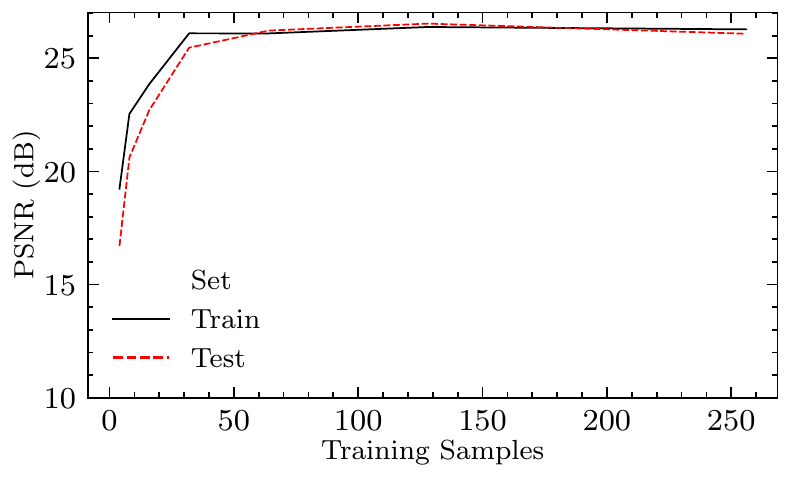}
  \caption{
    An MS-D network was trained on subsets of the data.
    The PSNR on the training set (black) and test set (remaining data;
    red) are displayed.
  }\label{fig:network-subsets}
\end{figure}

  \textbf{Generalization} We tested whether the network could be trained on
  fewer data samples and generalize to unseen data.  We used the
  1024-angle foam dataset, the MS-D network, 4 splits, and the X:1
  strategy.  The network was trained on the first 4, 8, 16, 32, 64, 128,
  and 256 slices of the data.  We report PSNR metrics on this
  \emph{training set} and on the remaining slices, which we refer to
  as the \emph{test set}.  The number of epochs was corrected for the
  smaller dataset size, such that all networks were trained for the
  same number of iterations.  When the training set exceeds 32 slices,
  the PSNR on the training and test set is comparable, as can be seen
  in Figure~\ref{fig:network-subsets}.

\section{Discussion}\label{sec:discussion}

The results show that the proposed Noise2Inverse method outperforms
conventional reconstruction algorithms SIRT and TV-MIN by a large
margin as measured in PSNR and SSIM\@.
This improvement is accomplished despite optimizing the
hyper-parameters of SIRT and TV-MIN on the clean reconstruction and
without likewise optimizing the Noise2Inverse hyper-parameters.
In addition, Noise2Inverse is able to significantly reduce noise in
challenging real-world experimental data, improving on the visual
impression obtained by SIRT and TV-MIN\@.

Extending the Noise2Self framework\cite{batson-2019-noise2}, we
describe a general framework for denoising linear image reconstructions
that provides a theoretical rationale for the success of
our method.
The framework shows that clean reconstructions may be recovered from
noisy measurements \emph{without observing clean measurements}, under
the common assumption that the measured noise is
element-wise independent and mean-zero.
  We remark that in low-noise situations, the trained network does not
  introduce additional artifacts in its output, as predicted by the
  theory.

  We now focus on the comparison between the proposed Noise2Inverse approach and
  the existing Noise2Noise and Noise2Self approaches.
  As in Noise2Noise, the network is presented with two noisy images during
  training.
  In Noise2Inverse, however, these images are sub-sampled reconstructions, and
  since the artifacts arising from sub-sampling the data are correlated, the
  input and target images are not statistically independent --- although the
  \emph{reconstructed noise} in these images is statistically independent.
  Therefore, our results fall outside of the Noise2Noise framework.
  As in Noise2Self, Noise2Inverse trains a denoiser from unpaired measurements.
  The key difference is that the noise is element-wise independent in the
  measurement domain, rather than in the reconstruction domain, where denoising
  takes place.
  Therefore, the results from~\cite{batson-2019-noise2} do not carry over to the
  inverse problems setting.
  However, we are able to prove Proposition~\ref{proposition:loss-decomposition}
  using essentially similar arguments to those in~\cite{batson-2019-noise2}.

The framework points the way to new applications of Noise2Inverse to
linear image reconstruction methods.
The implementation of Noise2Inverse for tomography shows that several
aspects are worth considering.
If reconstruction artifacts arise in the absence of noise, they will
be preserved.
In addition, if the reconstruction algorithm filters the noise at
the expense of resolution, this will cause blurring in the output
of our method.
Moreover, splitting the measurement uniformly can avoid biasing the
output of the method towards a particular subset of the measured data.
Finally, the performance of the neural network can be improved by
ensuring that the sub-reconstructions are homogeneously informative
throughout the image.

  Noise2Inverse is well-suited to imaging modalities that permit
  trading acquisition speed for measurement noise, as it aims to
  remove measurement noise but does not remove artifacts
  resulting from under-sampling,
  Whether this trade-off is possible, depends on the specifics of the
  imaging modality.
  Tomographic acquisition, for instance, permits acquiring the same
  number of projection images by lowering the exposure time at the
  cost of increased noise~\cite{mccollough-2017-low-dose}.
  Magnetic Resonance Imaging (MRI), on the other hand, is usually
  accelerated by reducing the number of measurements, rather than by
  acquiring noisier measurements~\cite{zbontar-2018-fastm}.
  Examples of imaging modalities that permit trading speed for noise
  include ultrasound imaging~\cite{matrone-2015-delay-multip}, deconvolution
  microscopy~\cite{sibarita-2005-decon-micros}, and X-ray holography~\cite{zabler-2005-optim-phase}.

The comparison of Noise2Inverse with Noise2Self demonstrates that the
success of our method depends not only on considerations of
statistical independence, but also on taking account of the physical
forward model.
Regarding statistical independence, we have demonstrated that a
straightforward application of Noise2Self fails on noisy tomographic
reconstructions due to coupling of the noise.
Regarding the forward model, we have investigated a two-step approach,
where Noise2Self is applied to projection or sinogram images --- which
\emph{do} satisfy the element-wise independence requirement --- before
reconstructing.
This approach performs worse than \mbox{TV-MIN} and Noise2Inverse in
terms of visual impression and quality metrics.
This matches earlier results~\cite{buchholz-2019-cryo-care}, and could
result from the fact that the consistency of the projection and
sinogram images with respect to the forward operator is not
necessarily preserved.
These results suggest that taking into account the properties of the
inverse problem --- as Noise2Inverse does --- significantly improves
the quality of the reconstruction.

Several variables affect the performance of Noise2Inverse.
Most importantly, the training strategy that reconstructs the input
images from at least as many projection angles as the target images
--- the X:1 strategy --- yields better results than vice versa.
This conclusion holds regardless of network architecture, number of
splits, or number of projection angles.
This suggests that noise in the gradient is less problematic than
noise in the input for neural network training, as was observed before~\cite{lehtinen-2018-noise2}.
Another variable that consistently predicts performance is the number
of angles; acquiring more projections yields a small but consistent
performance boost.
The number of parts in which the measured data is split, however,
deserves more nuance: when the projection angles are under-sampled,
the results indicate that two parts yield the best results; otherwise,
splitting into more parts yields better results.
Finally, maximal performance can be obtained by tuning the neural
network architecture and number of training iterations.
When tuning is not an option, an MS-D network can be trained with
limited risk of overfitting the noise.
  Finally, the object under study influences the comparative advantage of our
  method to conventional reconstruction techniques.
  When the aim is to retrieve low-contrast details from low-noise
  reconstructions, the difference may be minimal.
  When the object is self-similar and the noise has high intensity, on the other
  hand, our method can significantly outperform other methods.

\section{Conclusion}\label{sec:conclusion}

We have proposed Noise2Inverse, a CNN-based method for denoising
linear image reconstructions that does not require any
additional clean or noisy data beyond the acquired noisy dataset.
On tomographic reconstruction problems, it strongly outperforms both
standard reconstruction techniques such as Total-Variation
Minimization, and self-supervised image denoising-based techniques,
such as Noise2Self.
We also demonstrate that the method is able to significantly reduce
noise in challenging real-world experimental datasets.

\appendix
\begin{proof}
\textbf{[of Proposition~\ref{proposition:loss-decomposition}]}
First, expand the squared norm~\cite[Lemma 3.12]{rynne-2008-linear-funct-analy}
\begin{align*}
\norm*{h(\rnoisyC) - \rnoisyJ}^2
=& \norm*{h(\rnoisyC) - \rcleanJ + \rcleanJ  - \rnoisyJ}^2  \\
=& \norm*{h(\rnoisyC) - \rcleanJ}^2 + \norm*{\rcleanJ - \rnoisyJ}^2   \\
 &+ 2\inner{h(\rnoisyC) - \rcleanJ}{\rcleanJ - \rnoisyJ}.
\end{align*}

Let \(\xpar \in \X\), \(\pcleanpar= \A \xpar\), and \(\jpar \in \J\).
Then, from Equation~\eqref{eq:noise-zero-mean}, we obtain
\begin{align}
\expect[\mu]{\rnoisyJ \mid \xpar, \jpar}
&= \expect[\mu]{\Rec_\jpar \pnoisyrv_\jpar \mid \xpar, \jpar} \nonumber \\
&= \Rec_\jpar \, \expect[\xrv, \noise]{\pcleanpar_\jpar + \noise_\jpar \mid \xpar} \nonumber\\
&= \Rec_\jpar \, \pcleanpar_\jpar \nonumber\\
&= \rclean_{\jpar},
\label{eq:15}
\end{align}
where we use that \(\Rec_\jpar\) is linear.

The noisy random variables \(\rnoisyC\) and \(\rnoisyJ\) are independent
conditioned on \(\xpar\) and \(\jpar\), since domains of \(\Rec_{\jpar}\)
and \(\Rec_{\jpar^C}\) do not overlap, and the noise \(\noise\) is
element-wise statistically independent.
This independence condition allows us to interchange the order
of the expectation and inner product~\cite[Proposition 2.3]{eaton-2007-chapt-random-vector}, which
yields, using Equation~\eqref{eq:15},
\begin{align*}
\E&\left[\inner{h(\rnoisyC) - \rcleanJ}{\rcleanJ - \rnoisyJ} \mid \xpar, \jpar \right] \\
&= \inner{\expect{h(\rnoisyC) - \rcleanJ \mid \xpar, \jpar}}
         {\expect{\rcleanJ - \rnoisyJ \mid \xpar, \jpar}}  \\
&= \inner{\expect{h(\rnoisyC) - \rcleanJ \mid \xpar, \jpar}}
         {0}  \\
&= 0.
\end{align*}

Using the tower property of expectation, we obtain
\begin{align*}
\E_{\mu} &\norm*{h(\rnoisyC) - \rnoisyJ}^2 \\
&= \expect{
   \expect{\norm*{h(\rnoisyC) - \rnoisyJ}^2 \mid \xrv, \jrv}
   } \\
&= \expect{
   \expect{\norm*{h(\rnoisyC) - \rcleanJ}^2 + \norm*{\rcleanJ - \rnoisyJ}^2 \mid \xrv, \jrv}
   } \\
&= \E_{\mu} \norm*{h(\rnoisyC) - \rcleanJ}^2 + \E_{\mu} \norm*{\rcleanJ - \rnoisyJ}^2.
\end{align*}
\end{proof}

Similar proofs can be found in~\cite{batson-2019-noise2,adler-2018-deep-bayes-inver}.

\section*{Acknowledgment}
The authors acknowledge financial support from the Dutch Research
Council (NWO), project numbers 639.073.506 and 016.Veni.192.235.
The authors would like to express their appreciation for the
discussions with Nicola Vigano, Jan-Willem Buurlage, Sophia Coban, and Felix Lucka.
The authors have made use of the following additional software
packages to compute and visualize the experiments: Snakemake, Sacred,
Matplotlib, pandas, and scikit-image~\cite{koster-2012-snakem,greff-2017-sacred-infras,mckinney-2011-pandas,hunter-2007-matpl,walt-2014-scikit-image}.

\ifCLASSOPTIONcaptionsoff{}
  \newpage
\fi



\bibliographystyle{IEEEtran}
\bibliography{IEEEabrv,submission_ieee}


% \end{thebibliography}


\begin{IEEEbiography}
  [{\includegraphics[width=1in,height=1.25in,clip,keepaspectratio]{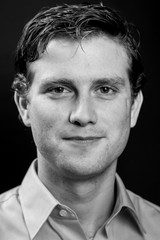}}]
  {Allard A. Hendriksen} received the
  M.Sc.\ degree in mathematics from the University of Leiden, Leiden,
  The Netherlands, in 2017.
  He is currently pursuing a Ph.D. degree with the Computational
  Imaging group at CWI, the national research institute for
  mathematics and computer science in Amsterdam, The Netherlands,
  focusing on combining deep learning and tomographic reconstruction
  algorithms.
\end{IEEEbiography}

\begin{IEEEbiography}
  [{\includegraphics[width=1in,height=1.25in,clip,keepaspectratio]{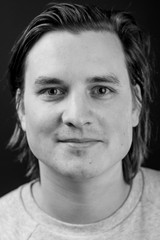}}]
  {Dani\"{e}l M. Pelt} received the M.Sc.\ degree in
  mathematics from the University of Utrecht, Utrecht, The Netherlands
  in 2010, and the Ph.D. degree at Leiden University, Leiden, The
  Netherlands, in 2016.
  As a Postdoctoral Researcher, he was at the Lawrence Berkeley
  National Laboratory (2016--2017), focusing on developing machine
  learning algorithms for imaging problems.
  He is currently a Postdoctoral Researcher with the CWI\@.
  His main research interest is developing machine learning
  algorithms for imaging problems, including tomographic \mbox{imaging}.
\end{IEEEbiography}

\begin{IEEEbiography}
  [{\includegraphics[width=1in,height=1.25in,clip,keepaspectratio]{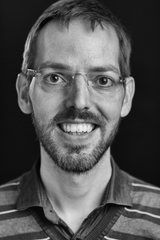}}]
  {K. Joost Batenburg} received the M.Sc.\ degree
  in mathematics and the M.Sc.\ degree in computer science from the
  University of Leiden, Leiden, The Netherlands, in 2002 and 2003,
  respectively, and the Ph.D. degree in mathematics in 2006. He
  currently leads the Computational Imaging group at Centrum Wiskunde
  \& Informatica, Amsterdam, The Netherlands. He is also professor of
  Imaging and Visualization at the University of Leiden.
\end{IEEEbiography}


%\vfill


\end{document}